# Modeling and analysis of the electromechanical behavior of surface-bonded piezoelectric actuators using finite element method

Huangchao Yu and Xiaodong Wang

***Abstract*—Piezoelectric actuators have been widely used to form a self-monitoring smart system to do SHM. One of the most fundamental issues in using actuators is to determine the actuation effects being transferred from the actuators to the host structure. This report summaries the state of the art of modeling techniques for piezoelectric actuators and provides a numerical analysis of the static and dynamic electromechanical behavior of piezoelectric actuators surface-bonded to an elastic medium under in-plane mechanical and electric loads using finite element method. Also case study is conducted to study the effect of material properties, bonding layer and loading frequency using static and harmonic analysis of ANSYS. Finally, stresses and displacements are determined, and singularity behavior at the tips of the actuator is proved. The results indicate that material properties, bonding layers and frequency have a significant influence on the stresses transferred to the host structure.**

## 1 INTRODUCTION

Piezoelectric actuators are quick in response, highly linear, small, non-invasive, inexpensive and easily wired into arrays. As a result, it has been widely used fault detection and structure health monitoring (SHM). The concept of using a network of piezoelectric actuators and sensors to form a self-monitoring and self-controlling smart system to do SHM in advanced structural design has drawn considerable interest among the research community [1-3]. In a reverse procedure of the piezoelectric sensors, an applied electric field to a piezoelectric actuator will result in a mechanical deformation of the actuator, which will in turn deform the host structure through load transfer at the interface. In these smart structures, both electromechanical coupling and material inhomogeneity are involved. The designers of such systems are constantly faced with the challenge of establishing suitable shapes and positions of actuators to provide high-performance structures.

One of the most fundamental issues in using integrated actuators in smart structures and SHM is to determine the actuation effects being transferred from the actuators to the host and the resulting overall structural response. Another important aspect related to the design of the integrated smart system is the determination of interfacial stresses that may result in failure of the structure integrity. Therefore, an accurate assessment of local stress and strain distribution would be really necessary in these smart structures involving the piezo-actuators and inhomogeneity. The subject of the modeling of the coupled electromechanical behavior of the surface-bonded piezoelectric actuators has received comprehensive attention from the scientific community. In the following part, the approaches that aim to achieve the coupled electromechanical behavior of the piezo-actuators bonded to the host structure are reviewed and summarized. These methods include both analytical, numerical and hybrid schemes.

### *1.1 Analytical approach*

There are mainly three kinds of analytical approaches to model the coupled electromechanical behavior of the piezo-actuators surface-bonded to the host structure, as show in the Table 1.

Table1. Comparison of three kinds of analytical approaches to model the coupled electromechanical behavior of the piezo-actuators

| Analytical models | References | Limitations |
|---|---|---|
| The shear-lag theory based on the Euler-Bernoulli model | [4-7] | ➢ the theory assumes linear strain distribution across the beam thickness, and this approximation only applies for low values of the frequency-thickness product of the lowest symmetric (S0) and anti-symmetric (A0) modes<br>➢ The theory cannot capture more than two lowest S0 and A0 modes with the increase of frequency. |
| The simplified pin-forced model | [8-11] | ➢ the model is a good approximation only if the Young's modulus and thickness of the actuator are small compared to those of the host structure or the bonding layer is very thin and stiff<br>➢ the model can only provide qualitative estimation about the actuation mechanism for low-frequency cases, which needs to be calibrated by either numerical simulation or experimental testing<br>➢ piezoelectric resonance effects cannot be captured in the model |
| The elasticity equation-based model | [12-13] | ➢ the model can provide the quantitative prediction of dynamic load transfer, but it is relatively complicated, only few references.<br>➢ In the model, the actuator thickness is assumed to be very small in comparison with its length, the applied electric filed primarily results in an axial deformation |

From the comparison, the elasticity equation-based model is most accurate one to consider the coupled electromechanical behavior between the piezoelectric actuator and the host medium, but it is too complicated to get the analytical solution. As the same, other two analytical approaches also have limitations. In order to make up and also verify the analytical models, numerical simulation techniques have been widely utilized to analyze the coupled electromechanical behavior of the piezo-actuators bonded to the host structure.

### *1.2 Numerical and hybrid approach*

In modeling the electromechanical interaction between the actuator and the host structure, some commercially available FE codes, e.g., COMSOL/Multi-physics and ANSYS, provide researchers convenient tools to conduct the coupled physical problem. FEM is a really powerful approach to model the behavior of piezo-actuators [14,15]. However, FE simulation lacks of the capability to provide a very clear physical explanation of the numerically predicted results.

Hybrid approaches provide potential solutions to compensate for the disadvantages of pure FE simulation. In the hybrid schemes, the FE solution using piezoelectric elements is only conducted in limited areas (e.g., the piezo-actuation area) to obtain the prescribed excitation, and then combined with analytical guided wave excitation model in the host structure. In the approach, the FE calculation is conducted to determine only the surface stresses or the volume forces created by the piezoelectric elements, which are used as the prescribed excitation for the analytical solution in the host medium [16,17].The hybrid schemes enable the calculation of piezoelectrically induced wave response in the infinite host medium with less computational effort, since the host structural model usually consumes much more elements than does the piezo-actuator model.[18]

### *1.3 Objective of the report*

Since the limitations and complication of analytical approaches and pure numerical methods, the hybrid approach is the best choice, which combines using FEM to determine the actuation effects being transferred from the actuators to the with wave propagation analytical solution in the host medium. The objective of the present report is to provide a comprehensive numerical study of the static and dynamic electromechanical behavior of piezoelectric actuators surface-bonded to an elastic medium under in-plane mechanical and electric loads, like the interface stresses transferred to host structure. Also case study will be conducted to study the effect of material properties, bonding layer and loading frequency upon the actuation process using static and harmonic analysis of ANSYS.

## 2 MODELING OF THE PROBLEM

### *2.1 Physical FEM introduction*

In order to model the behavior of piezoelectric actuators, physical FEM will be used, as shown in Fig.1. In this modeling part, the physical system, idealization and discretization will be explained.

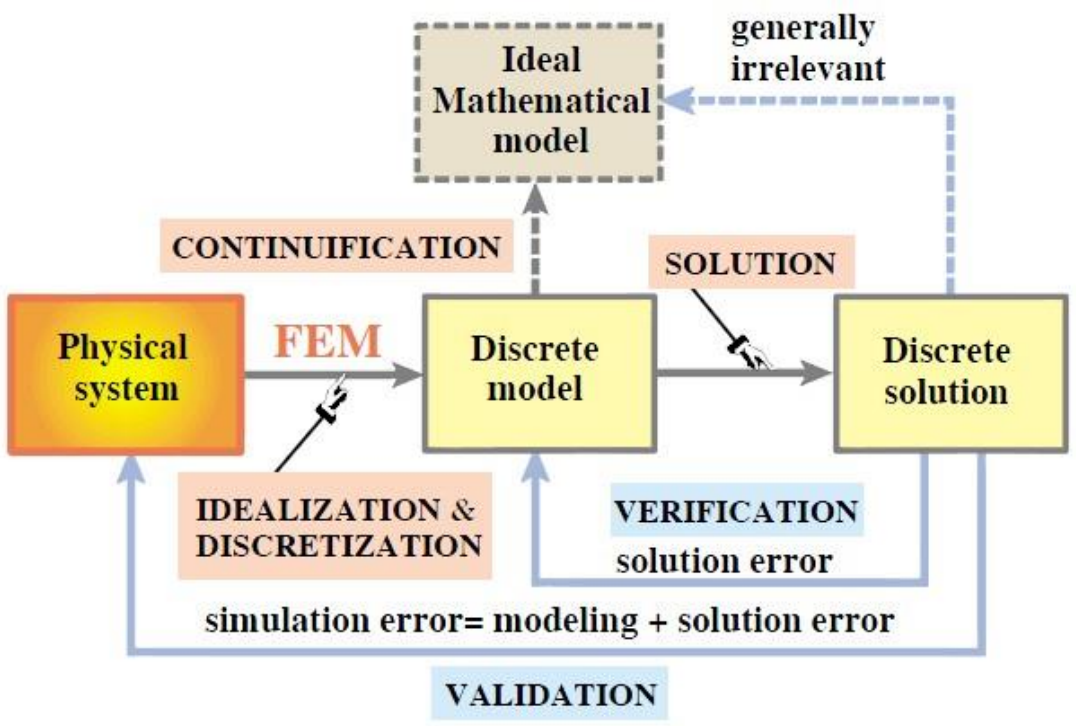

Fig.1 The Physical FEM Process Diagram (Taken from [18])

### *2.2 Physical system*

The concept of using a network of piezoelectric actuators and sensors to form a self-monitoring and self-controlling smart system to do structure health monitoring in advanced structure has been applied in the industry, for example, aerospace, aircraft and marine structures.

For example, a flight is considered as the original physical system, on whose surface a network of piezoelectric actuators and sensors are bonded to get the health and fault information of aircraft skin.

### *2.3 Idealization*

In order to simplify the problem, we just need to study the behavior of one of the piezoelectric actuators first, because every unit is the same. The actuators used in the system is used to generate diagnosis wave, so assumption is given that such kind of actuator is choose here which its length is much larger than its width and height. Based on this assumption, the model is idealized as Fig.2. A piezoelectric actuator is surface-bonded to a very large host structure.

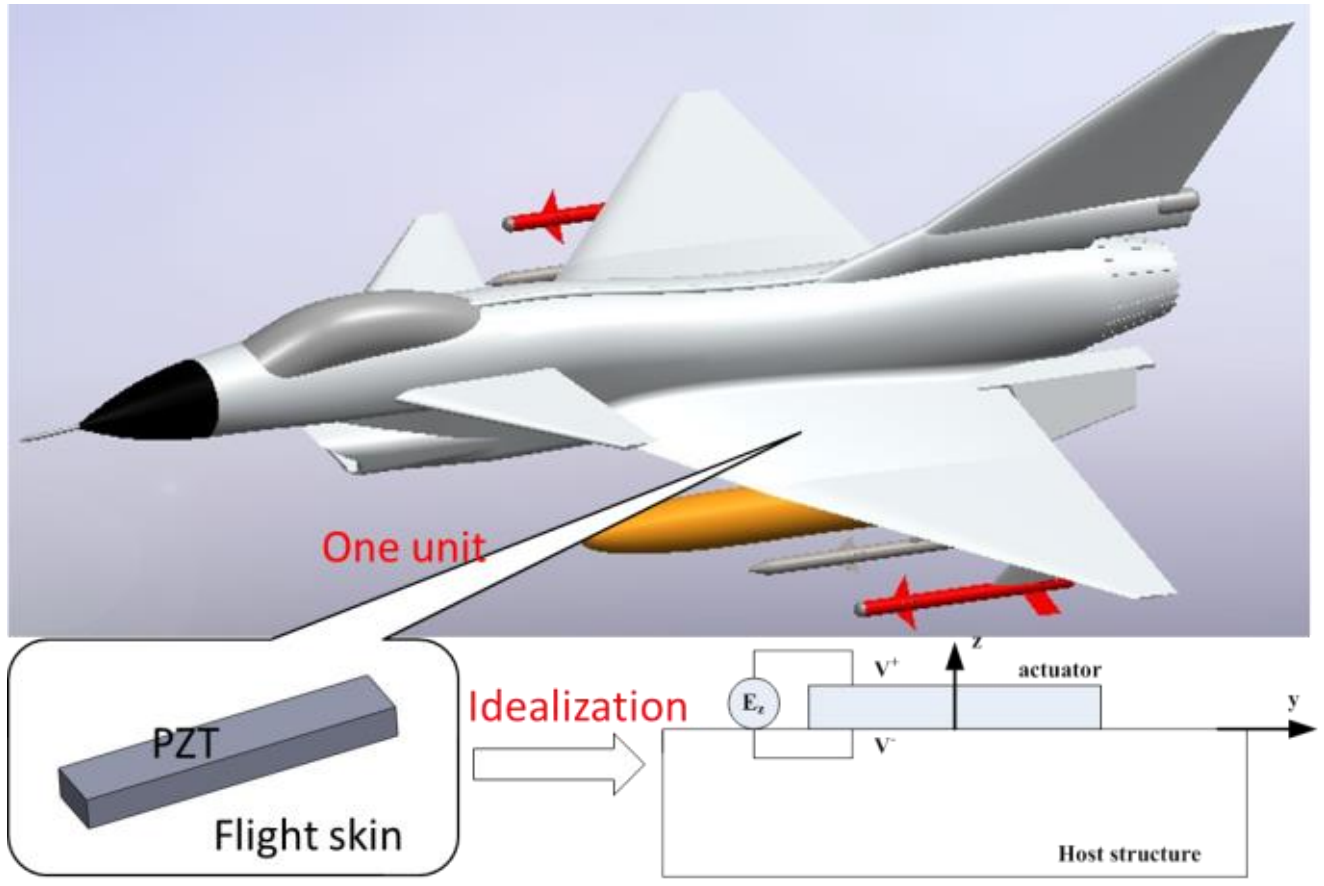

Fig.2 Physical system and idealization process (Model is built in SolidWorks)

According to the results of [12-13] and midterm project report, the stresses transferred to the host structure and boundary conditions are symmetric. So the model can be modified as shown in Fig.3.

Corresponding author: huangcha@ualberta.ca, Department of Mechanic Engineering, University of Alberta, Canada

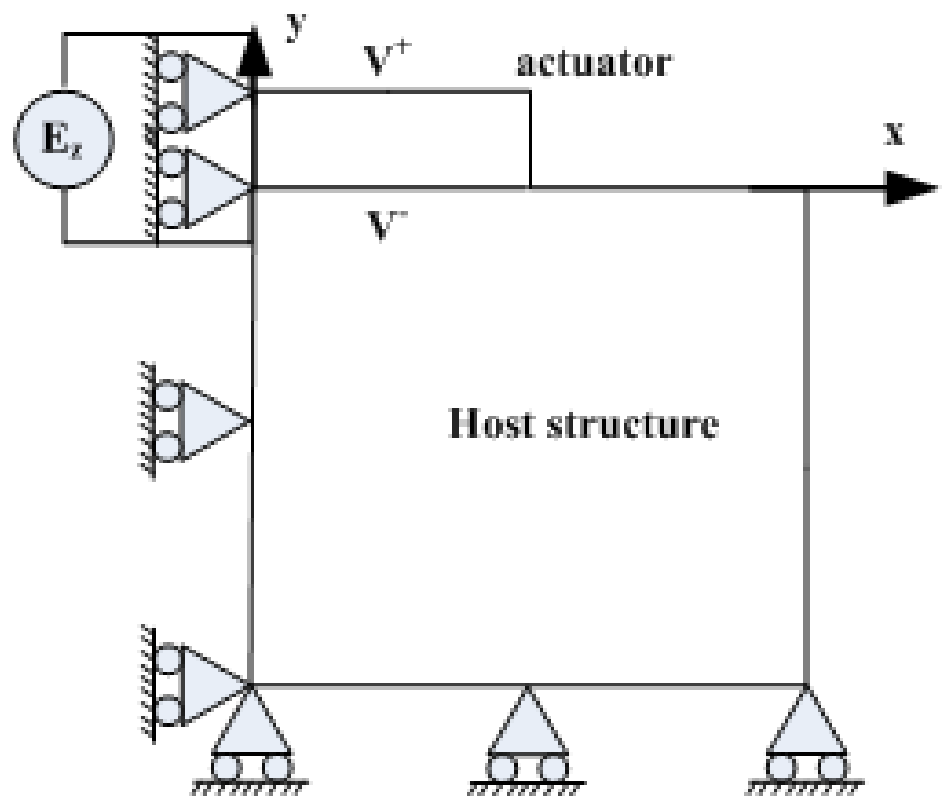

Fig.3 modified symmetric model including loads and boundary conditions

*2.4 Discretization*

After the idealization of the problem, discretization is conducted to get the discrete model. The mesh method here is so-called Global-local analysis [19]. In the global stage the behavior of the entire structure is simulated with a finite element model that necessarily ignores details such as cutouts or joints. These details do not affect the overall behavior of the structure. On the other hand, for some local parts, a largely regular mesh should be used to get the detail local information.

For the host medium in this project, the mesh principle is using an accurate mesh near the actuator and the coarser the further. The mesh result is shown in the following Fig.4

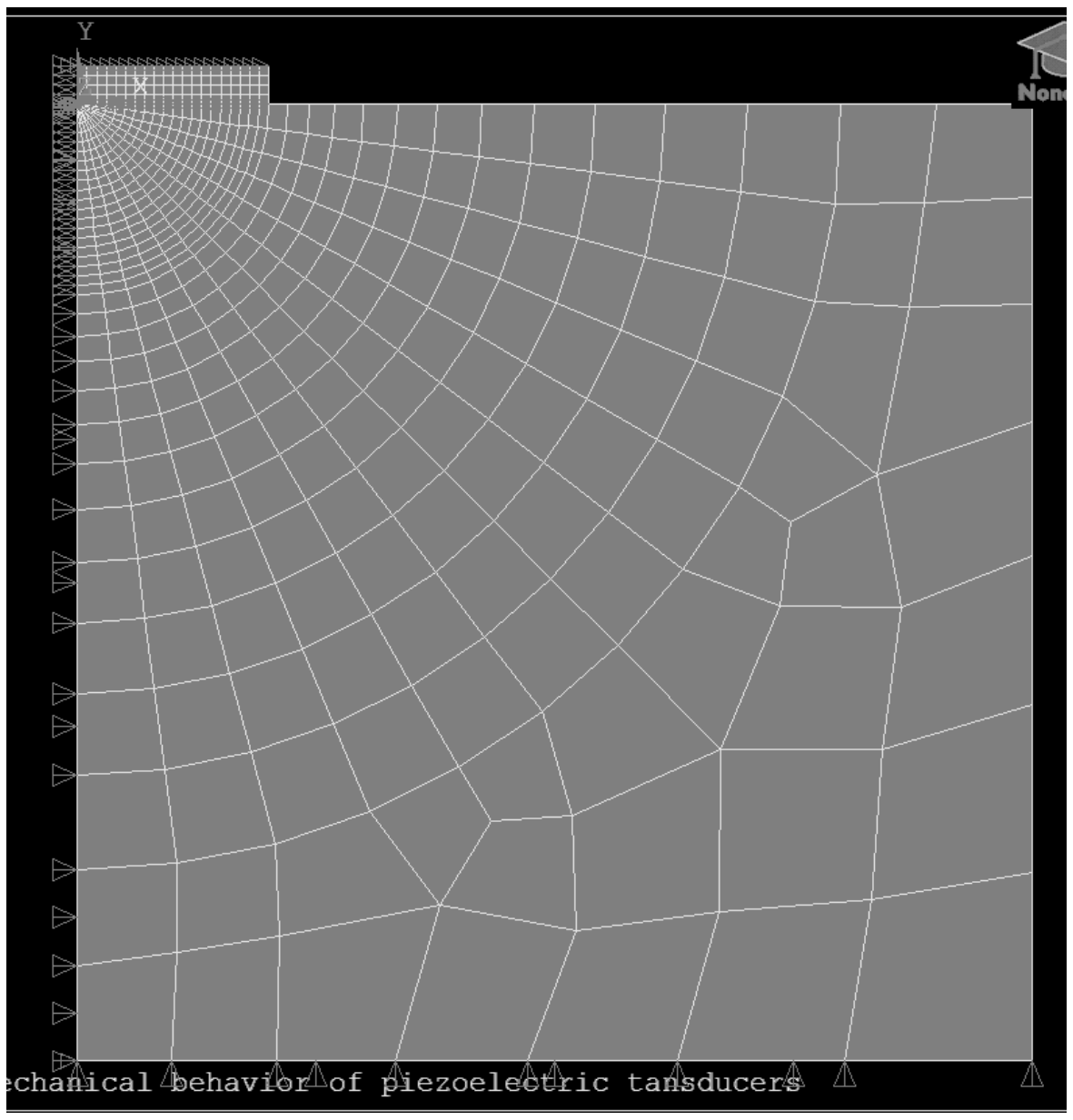

Fig.4 Mesh result of the model using global-local analysis method

Now if element type, material properties, geometry and boundary conditions are given, we can get the solution of stress and strain distribution near the actuator using ANSYS.

## 3 Analysis, Results and discussion

*3.1 ANSYS simulation of a case study*

A case study is simulated in ANSYS, which uses the PZT-4 piezoelectric actuator. To simulate the case which the host medium is infinite, a piezoelectric actuator with length=10mm and height=1mm surface-bonded to a matrix of 100mm*50mm was considered. In the preprocessor part, the following settings are used: element type is coupled-field solid plane 13 for piezoelectric actuators, and Solid Plane 182 for the host medium; material properties are shown in Table 2 and 3. In order to compare with the analytical results, these parameters are the same to reference [20]. Also the same to [20], **plain strain** condition is assumed [the actuator studied here: its length is much larger than its width and height.]

Table2. Material properties and geometry of the piezoelectric actuator

| Elastic stiffness parameters | $c_{11}$ | $c_{12}$ | $c_{13}$ | $c_{33}$ | $c_{44}$ |
|---|---|---|---|---|---|
| $(\times 10^{10} Pa)$ | 13.9 | 6.78 | 7.43 | 11.5 | 2.56 |
| Piezoelectric constants | $e_{31}$ | $e_{33}$ | $e_{15}$ | | |
| $(C/m^2)$ | -5.2 | 15.1 | 12.7 | | |
| Dielectric constants | $\varepsilon_{11}$ | $\varepsilon_{33}$ | | | |
| $\times 10^{-9} C/Vm$ | 6.45 | 5.62 | | | |
| Geometry | a | h | | | |
| m | 0.01 | 0.001 | | | |

Table3. Material properties of the host medium

| | |
|---|---|
| Young's Modulus $(\times 10^{10} Pa)$ | 5.27 |
| Poisson's ratio | 0.3 |

*3.2 Static analysis and result*

*3.2.1 Basic model result*

After applied the boundary conditions, that is fixing all the DOFs of the bottom surface of the host medium, and applying voltage 100 and 0 on the upper and lower surface of the actuator, we can get the solutions and results as shown in Fig.5-10.

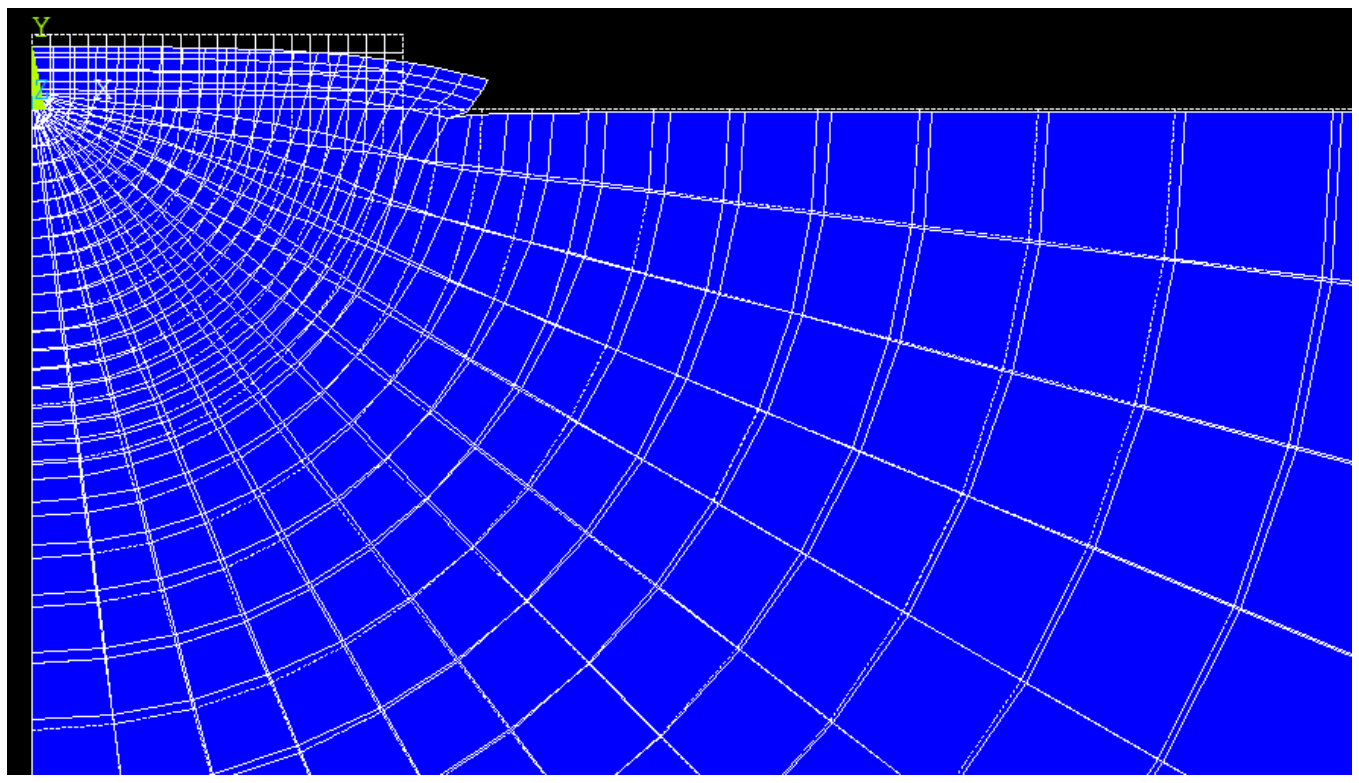

Fig.5 deformed shape of static analysis

Corresponding author: huangcha@ualberta.ca, Department of Mechanic Engineering, University of Alberta, Canada

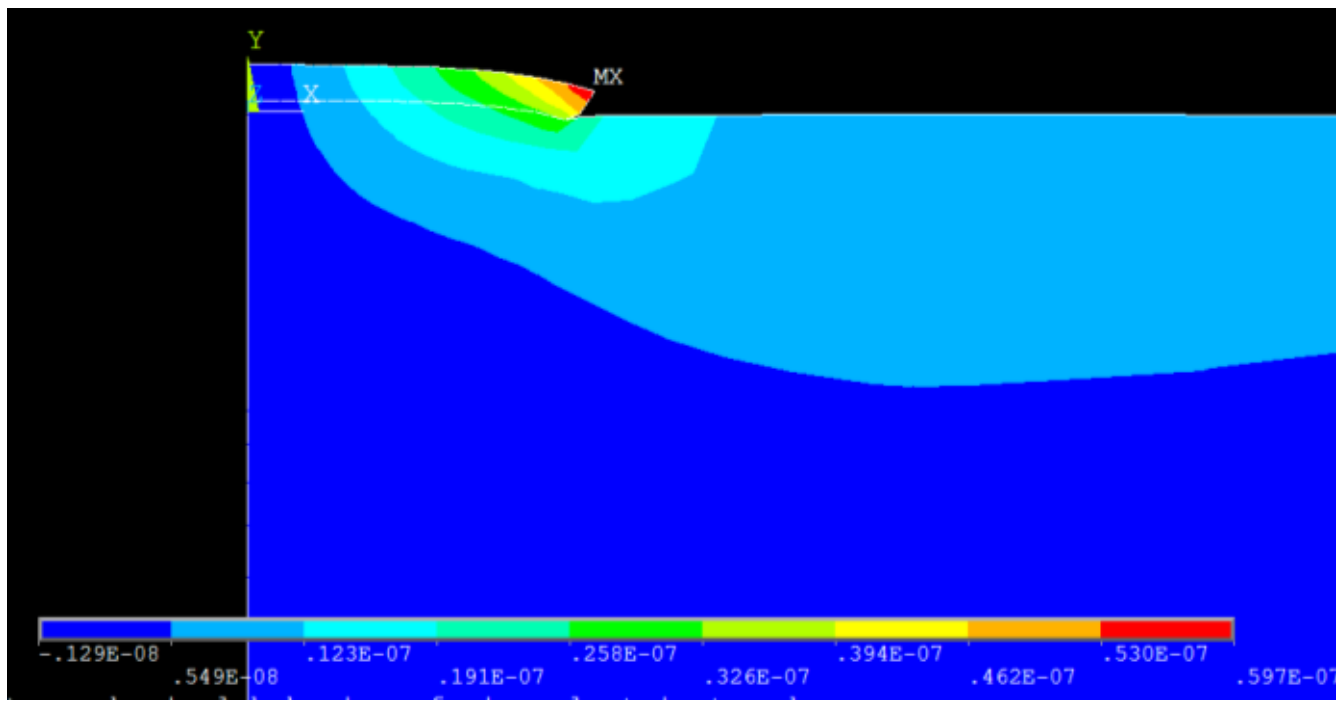

Fig.6 X- displacement of static analysis (nodal solution)

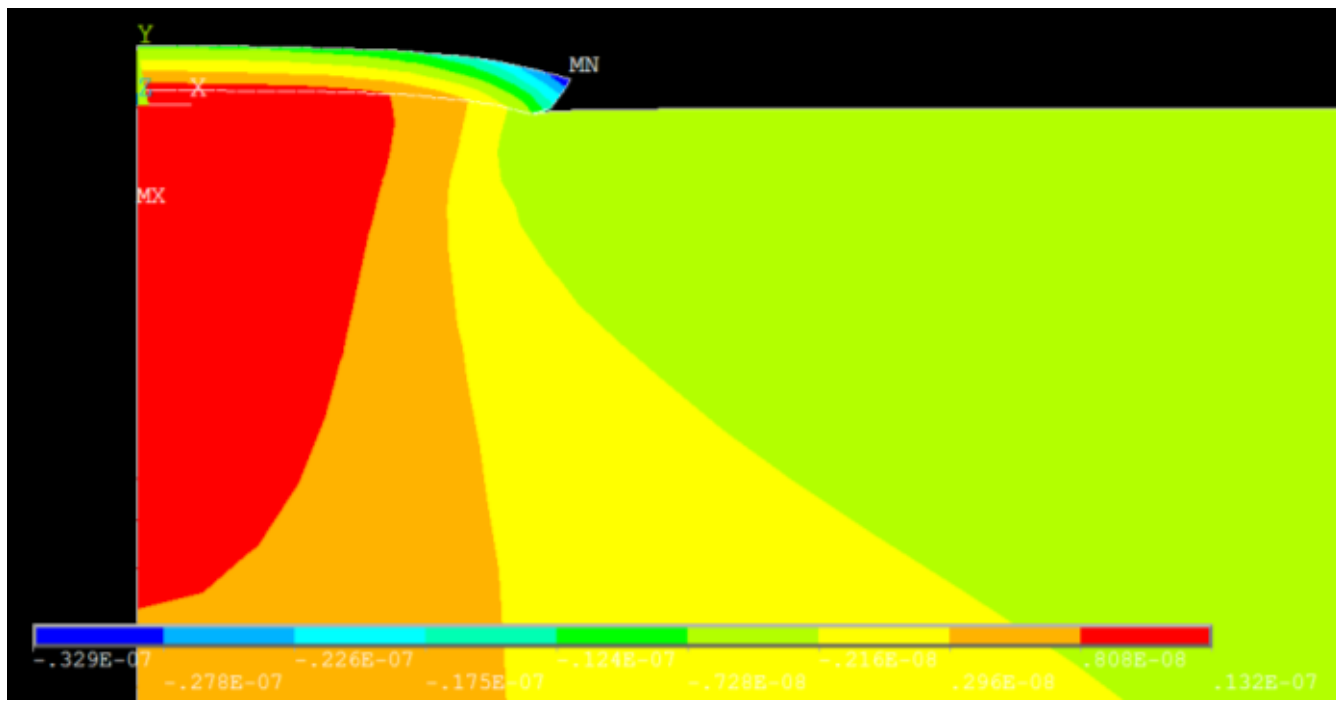

Fig.7 Y-displacement of static analysis (nodal solution)

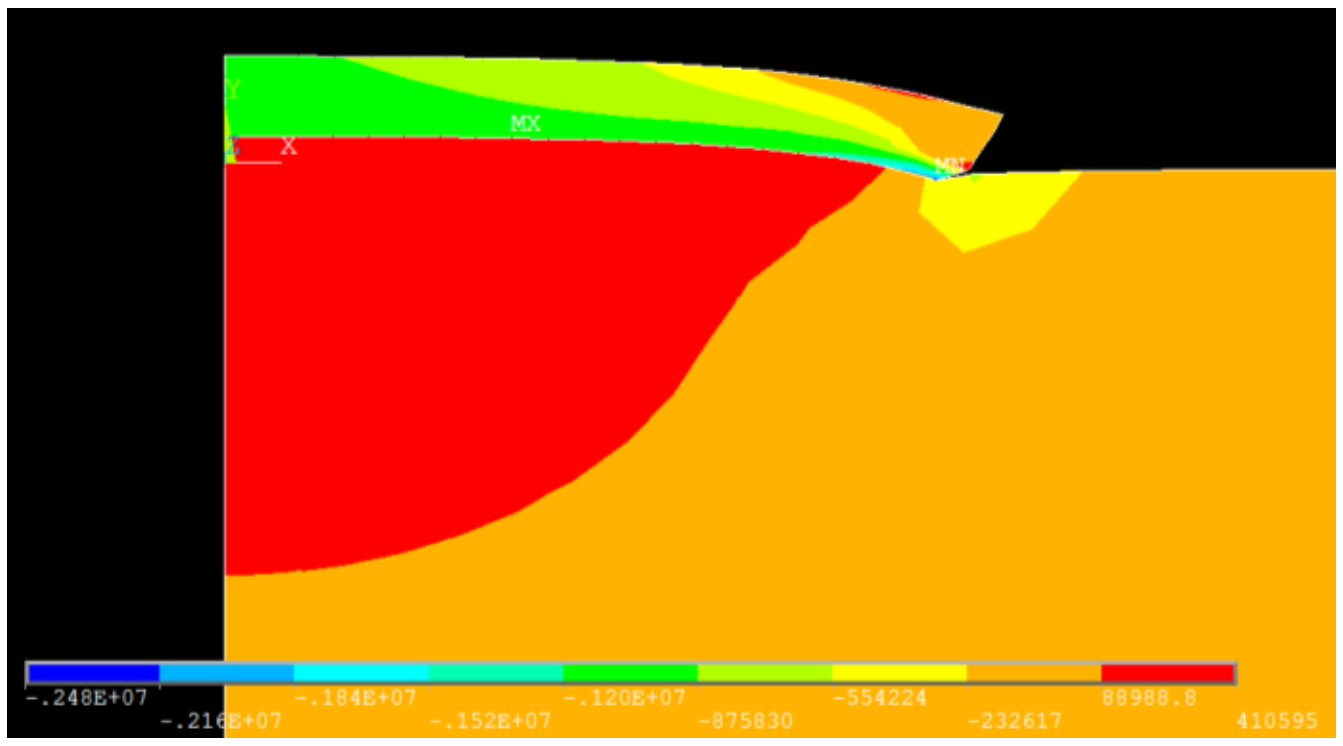

Fig.8 X-stress of static analysis (nodal solution)

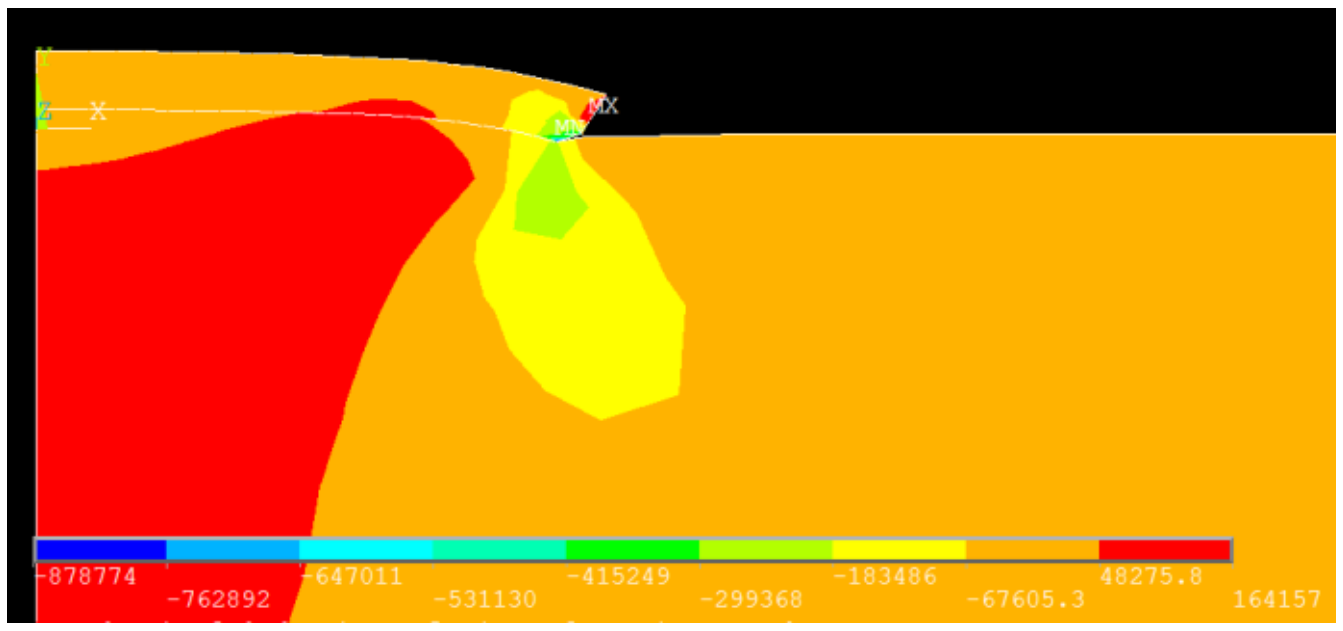

Fig.9 Y-stress of static analysis (nodal solution)

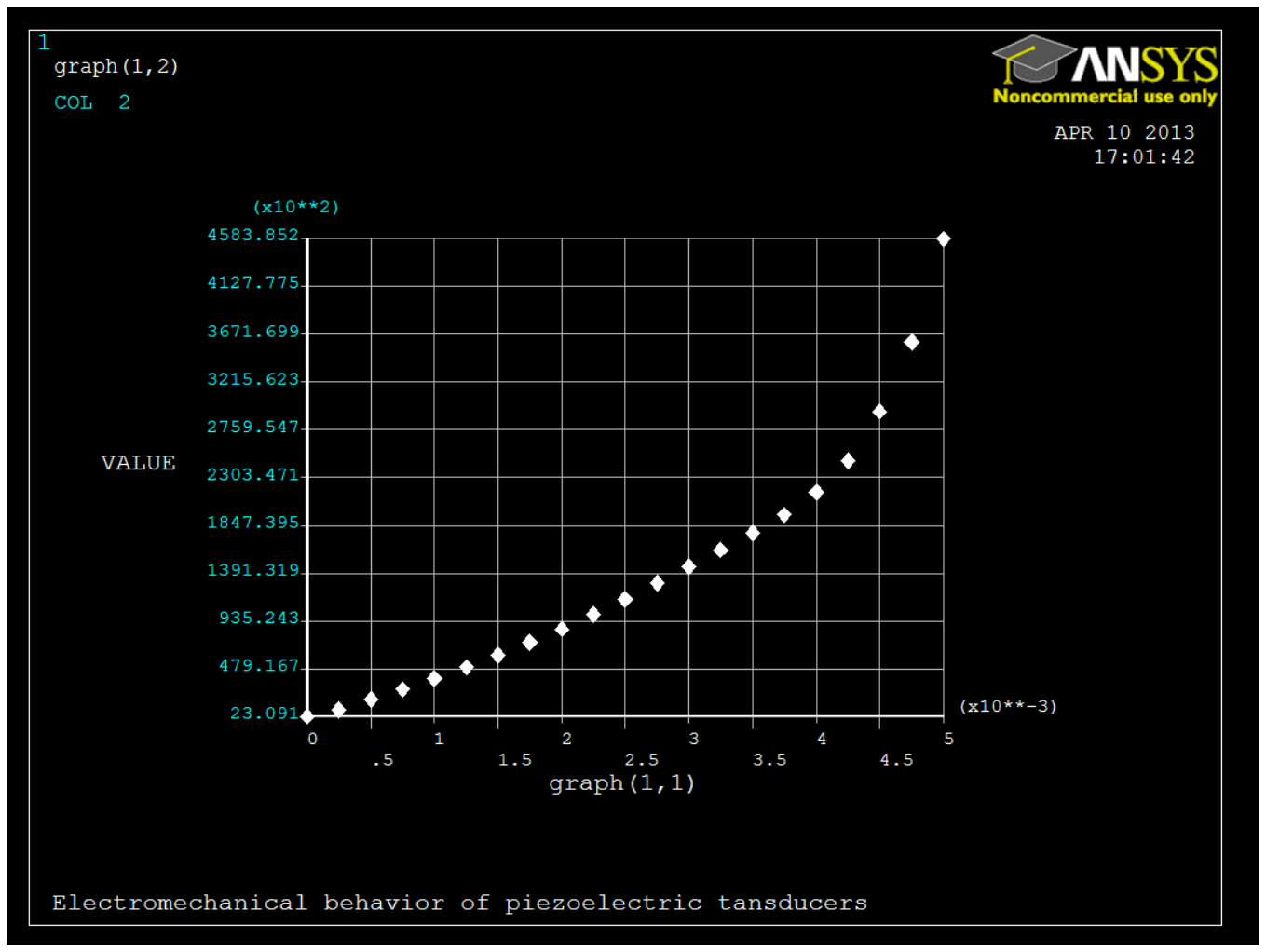

Fig.10 XY shear stress at interface (graph(1,1) is the interface nodes position---meter, graph(1,2 is the shear stress---Pa)

Displacement and stress transferred from the actuator to the host are determined. By comparing the displacement and stress result in X and Y direction, we find displacement and stress in X direction are much larger than that in Y direction, which proves that the assumption of the elasticity equation-based model. From Fig.10, the XY shear stress transferred to the host by the actuator is determined and there is a singularity at the tips of the actuator.

### *3.2.2 Comparison of nodal and element solution*

In FEM, stresses and strains are calculated at the elements, so the element solution is accurate result of the calculation, but it is discontinuous across elements. On the other hand, each node will have multiple values from each element it is attached to. The averaged stress/strain value is considered as the nodal solution which is continuous across elements. By checking the difference of nodal solution and element solution, as shown in Fig.11-12, we can conclude that the mesh used in the project is good, because these two result almost the same.

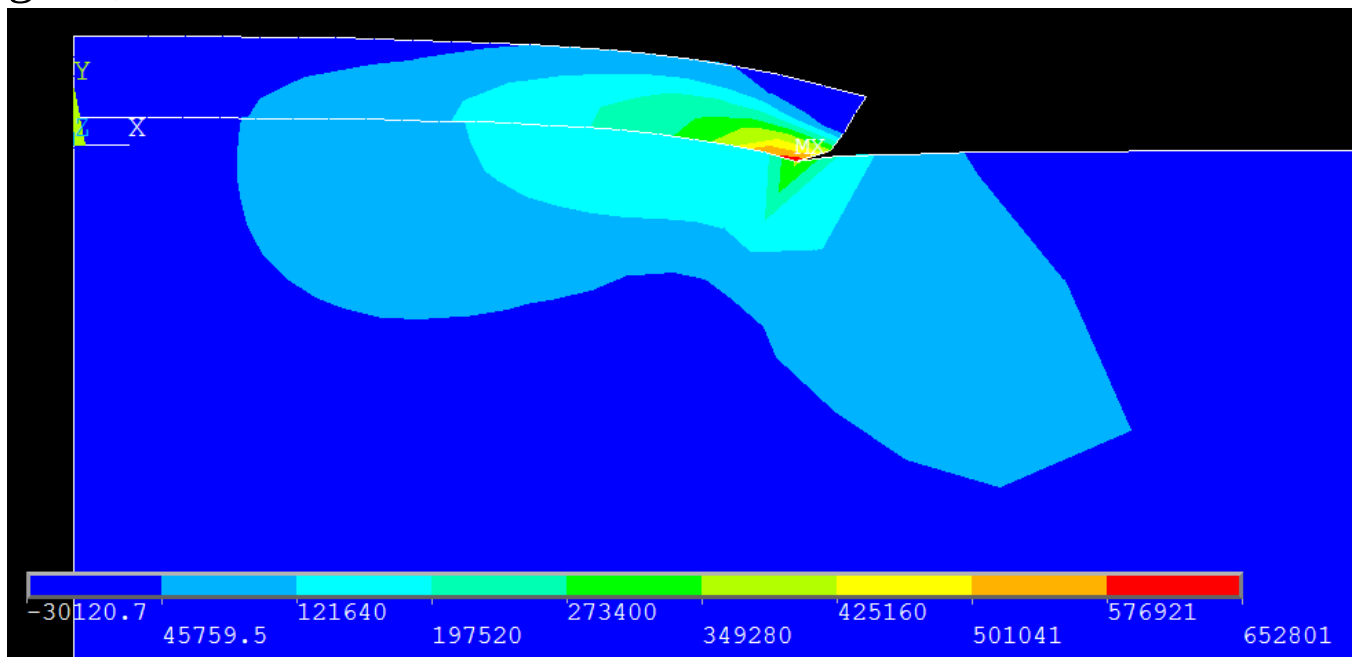

Fig.11 Nodal solution of the XY shear stress around the actuator

Corresponding author: huangcha@ualberta.ca, Department of Mechanic Engineering, University of Alberta, Canada

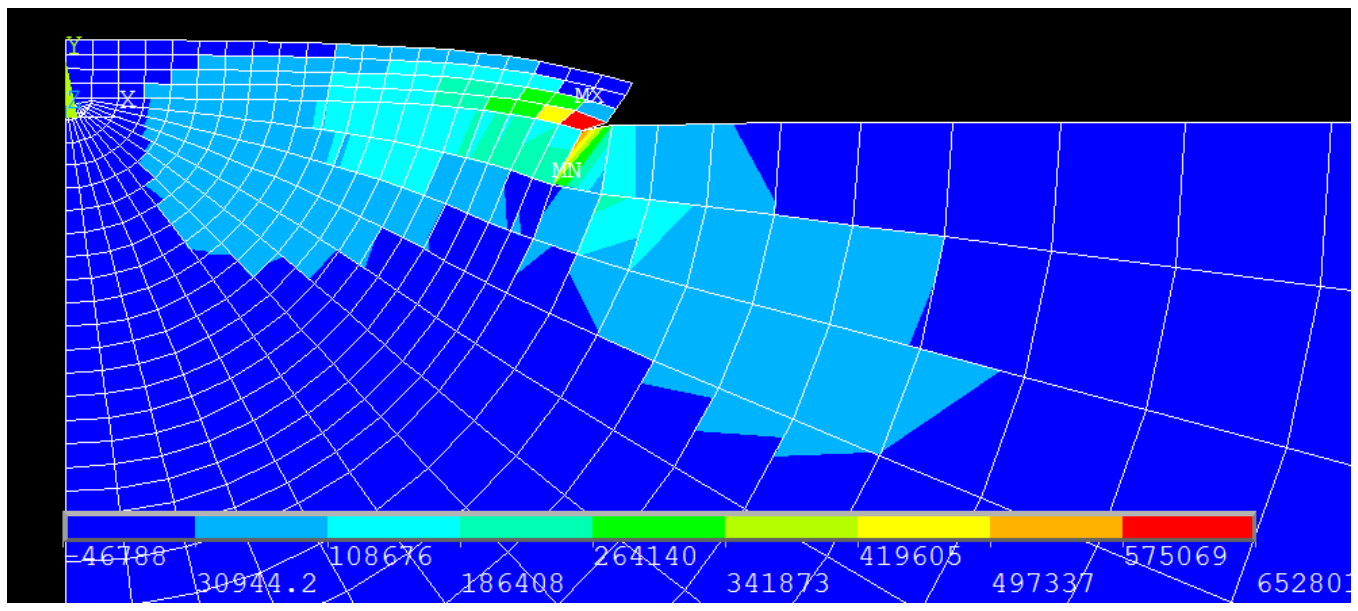

Fig.12 Element solution of the XY shear stress around the actuator

### *3.2.3 Comparison with analytical model*

XY shear stress result of the nodes at interface can be listed as shown in table 4, and these data of the shear stress along the interface can be plotted in Fig.13.

Table4. XY shear stress ( $N / m^2$ ) along the interface

| x/a | NODE | SX | SY | SXY |
|---|---|---|---|---|
| 0.05 | 2 | -1021900 | 19046 | 8892 |
| 0.1 | 21 | -1021500 | 19698 | 18156 |
| 0.15 | 20 | -1020000 | 20781 | 27727 |
| 0.2 | 19 | -1018300 | 22366 | 37705 |
| 0.25 | 18 | -1016400 | 24438 | 48375 |
| 0.3 | 17 | -1015000 | 27267 | 59846 |
| 0.35 | 16 | -1014200 | 30749 | 72132 |
| 0.4 | 15 | -1014800 | 34894 | 85393 |
| 0.45 | 14 | -1017300 | 39478 | 100050 |
| 0.5 | 13 | -1023300 | 45012 | 116160 |
| 0.55 | 12 | -1033300 | 50715 | 133670 |
| 0.6 | 11 | -1049500 | 56493 | 152410 |
| 0.65 | 10 | -1072500 | 60621 | 172320 |
| 0.7 | 9 | -1105100 | 61230 | 193760 |
| 0.75 | 8 | -1149300 | 54679 | 218460 |
| 0.8 | 7 | -1215000 | 32480 | 252300 |
| 0.85 | 6 | -1327400 | 3326.5 | 302770 |
| 0.9 | 5 | -1558800 | -91799 | 400370 |
| 0.95 | 4 | -2101400 | -79717 | 560350 |

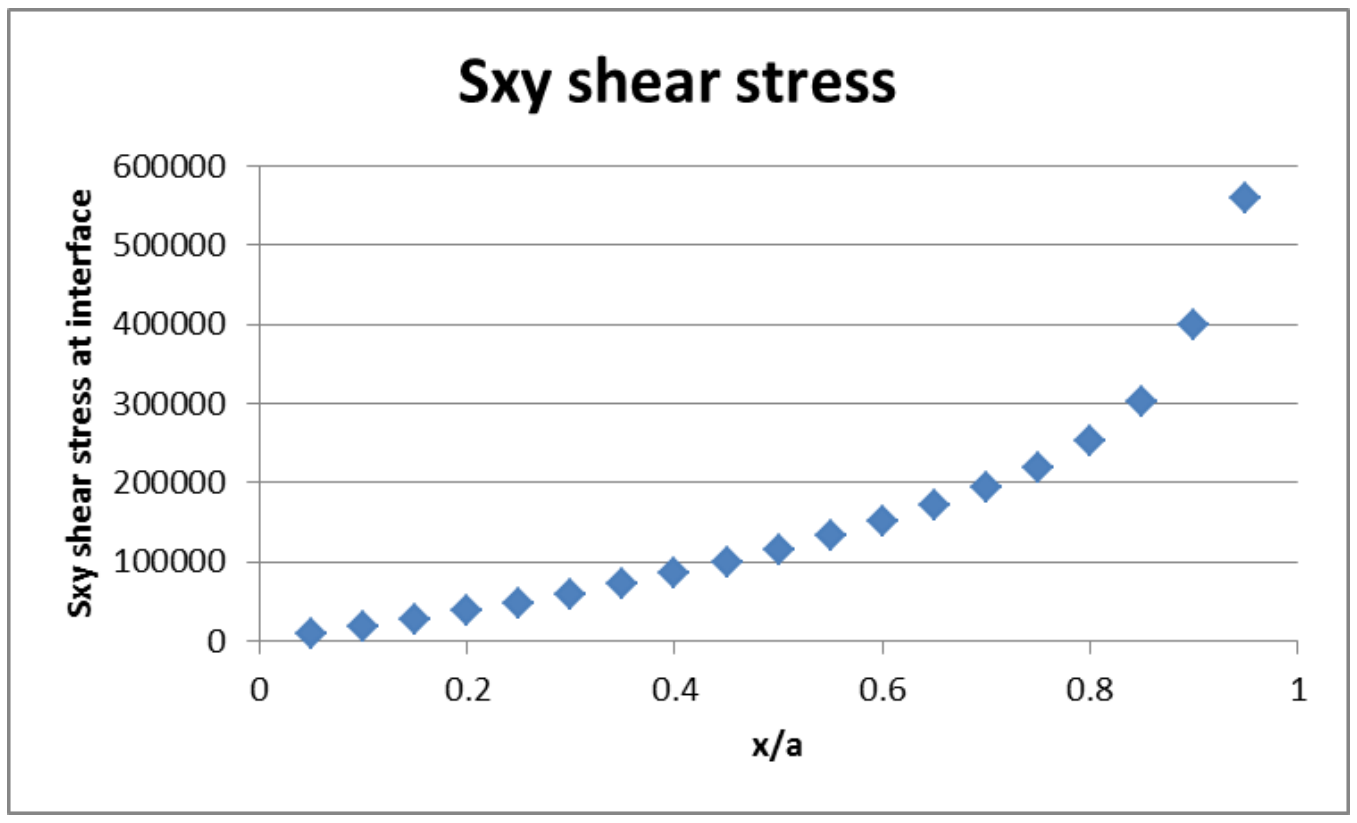

Fig.13 the shear stress distribution along the interface

Comparing the result with that of the analytical model in reference [20], as shown in Fig.14, singularity behavior of piezoelectric actuators is shown, the actuation shear being transferred from the actuator to the host is determined and the finite element method used in this project is verified.

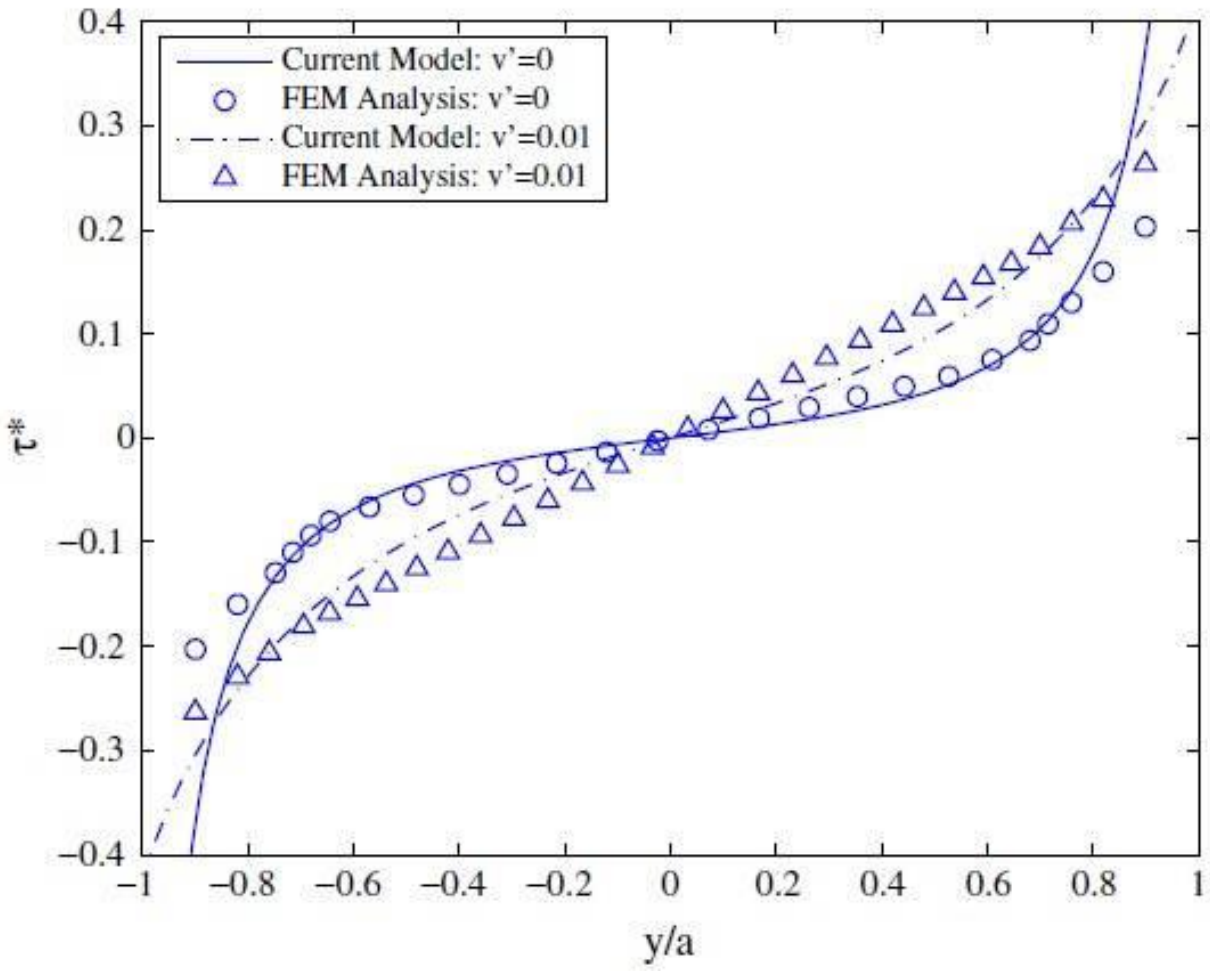

Fig.14 the shear distribution along the interface determined by the FEM and the analytical model in [20].

### *3.2.4 Effect of material properties*

One of the key issues to use the actuator to do structure health monitoring is how to optimize the actuation effects and how to get the resulting overall structural response we need. In this part, the effects of material properties will be studied.

The material properties include Young's modulus and Poisson's ratio of the structure, elastic stiffness, piezoelectric constants, dielectric constants of the actuator.

For example, in the previous static basic model , the Young's modulus of the structure is 5.27GPa, here in order to get an clear and large enough difference between the solutions, ten times Young's modulus is concerned, i.e. E=52.7GPa.

After applied the boundary conditions, that is fixing all the DOFs of the bottom surface of the host medium, and applying voltage 100 and 0 on the upper and lower surface of the actuator, we can get the solutions and results as shown in Fig.15-17

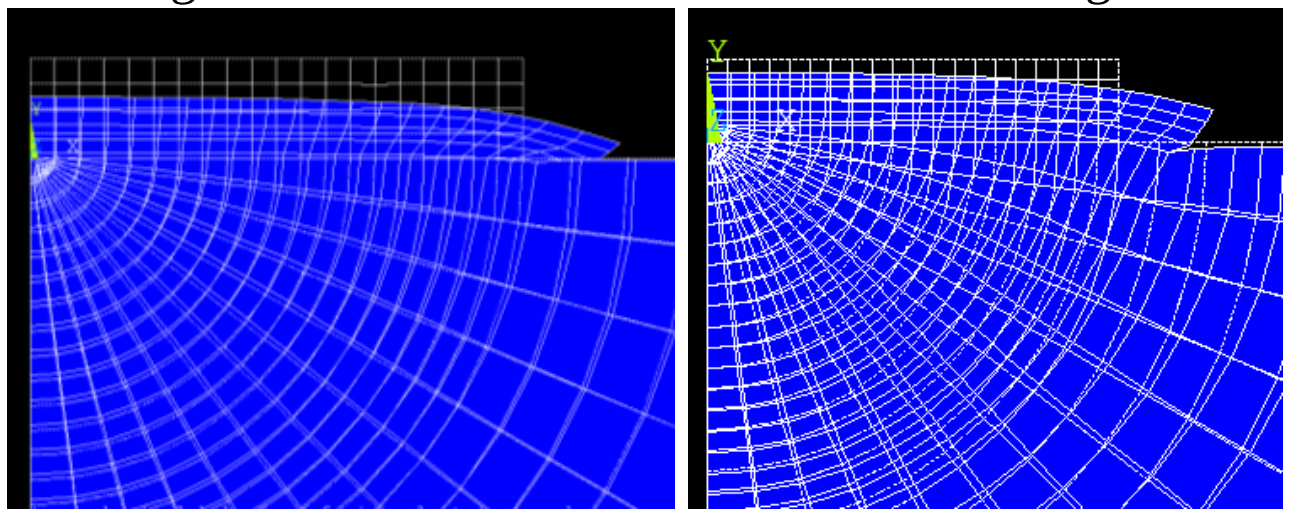
Fig.15 deformed shape comparison (left: E=52.7GPa, right: E=5.27GPa)

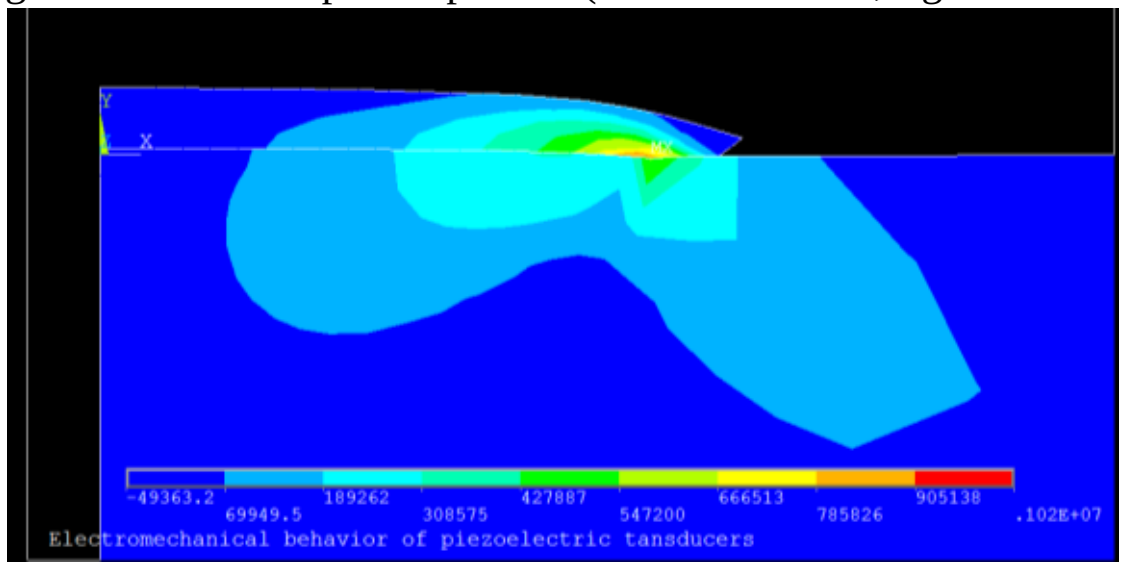

Fig.16 XY shear stress---nodal solution (E=52.7Gpa)

Corresponding author: huangcha@ualberta.ca, Department of Mechanic Engineering, University of Alberta, Canada

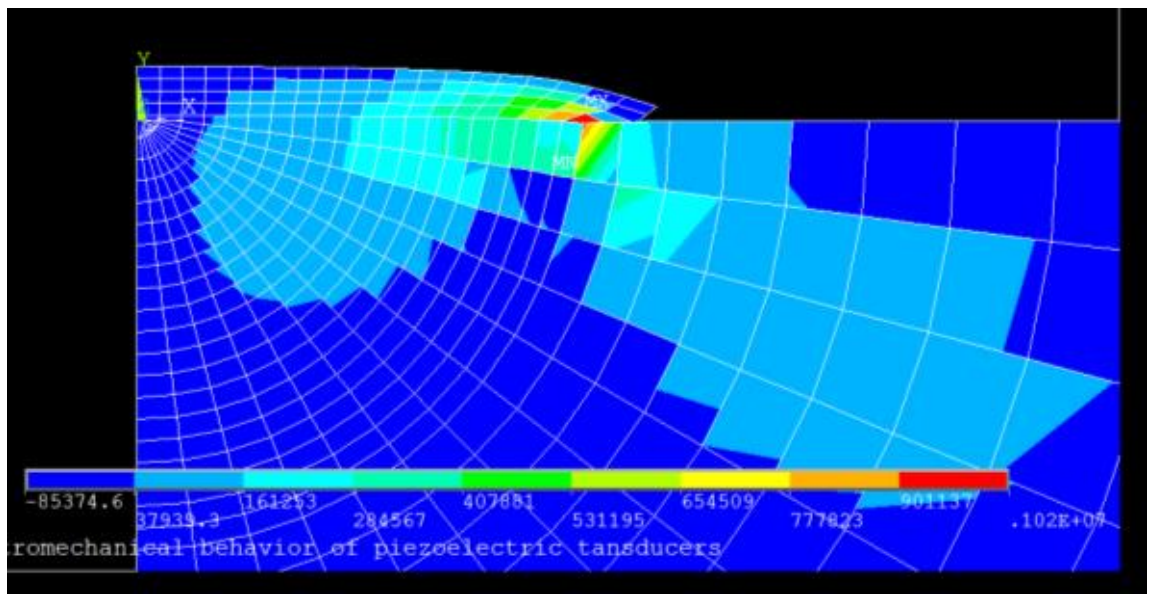

Fig.17 XY shear stress---element solution (E=52.7Gpa)

As shown in figure 16 and 17, the nodal and element solution of the XY shear stress are almost the same, which means the mesh is good enough in this analysis. By comparing the deformed shape in Fig.15, the deformation in the larger Young's modulus model is much larger than that in the basic model. It shows that the Young's modulus has a significant influence on the actuator effect. In details, the shear stress result at the interface transferred to host can be list in the following table. Then the XY shear stress at interface with material properties E=5.27GPa and 10E can be plotted to compare, as shown in Fig.18.

Table5. Stress result at interface (E=52.7Gpa)

| x/a | NODE | SX | SY | SXY |
|---|---|---|---|---|
| 0.05 | 2 | -1.55E+06 | 19892 | 9584.5 |
| 0.1 | 21 | -1.55E+06 | 20871 | 19559 |
| 0.15 | 20 | -1.55E+06 | 22596 | 30187 |
| 0.2 | 19 | -1.55E+06 | 25088 | 41802 |
| 0.25 | 18 | -1.55E+06 | 28488 | 54911 |
| 0.3 | 17 | -1.54E+06 | 33168 | 69909 |
| 0.35 | 16 | -1.54E+06 | 39249 | 87133 |
| 0.4 | 15 | -1.54E+06 | 47017 | 1.07E+05 |
| 0.45 | 14 | -1.53E+06 | 56748 | 1.31E+05 |
| 0.5 | 13 | -1.53E+06 | 69480 | 1.59E+05 |
| 0.55 | 12 | -1.52E+06 | 85074 | 1.91E+05 |
| 0.6 | 11 | -1.52E+06 | 1.04E+05 | 2.28E+05 |
| 0.65 | 10 | -1.52E+06 | 1.24E+05 | 2.71E+05 |
| 0.7 | 9 | -1.54E+06 | 1.41E+05 | 3.19E+05 |
| 0.75 | 8 | -1.56E+06 | 1.48E+05 | 3.74E+05 |
| 0.8 | 7 | -1.62E+06 | 1.31E+05 | 4.44E+05 |
| 0.85 | 6 | -1.75E+06 | 84877 | 5.36E+05 |
| 0.9 | 5 | -1.98E+06 | -56292 | 6.97E+05 |
| 0.95 | 4 | -2.64E+06 | -2.04E+05 | 9.16E+05 |

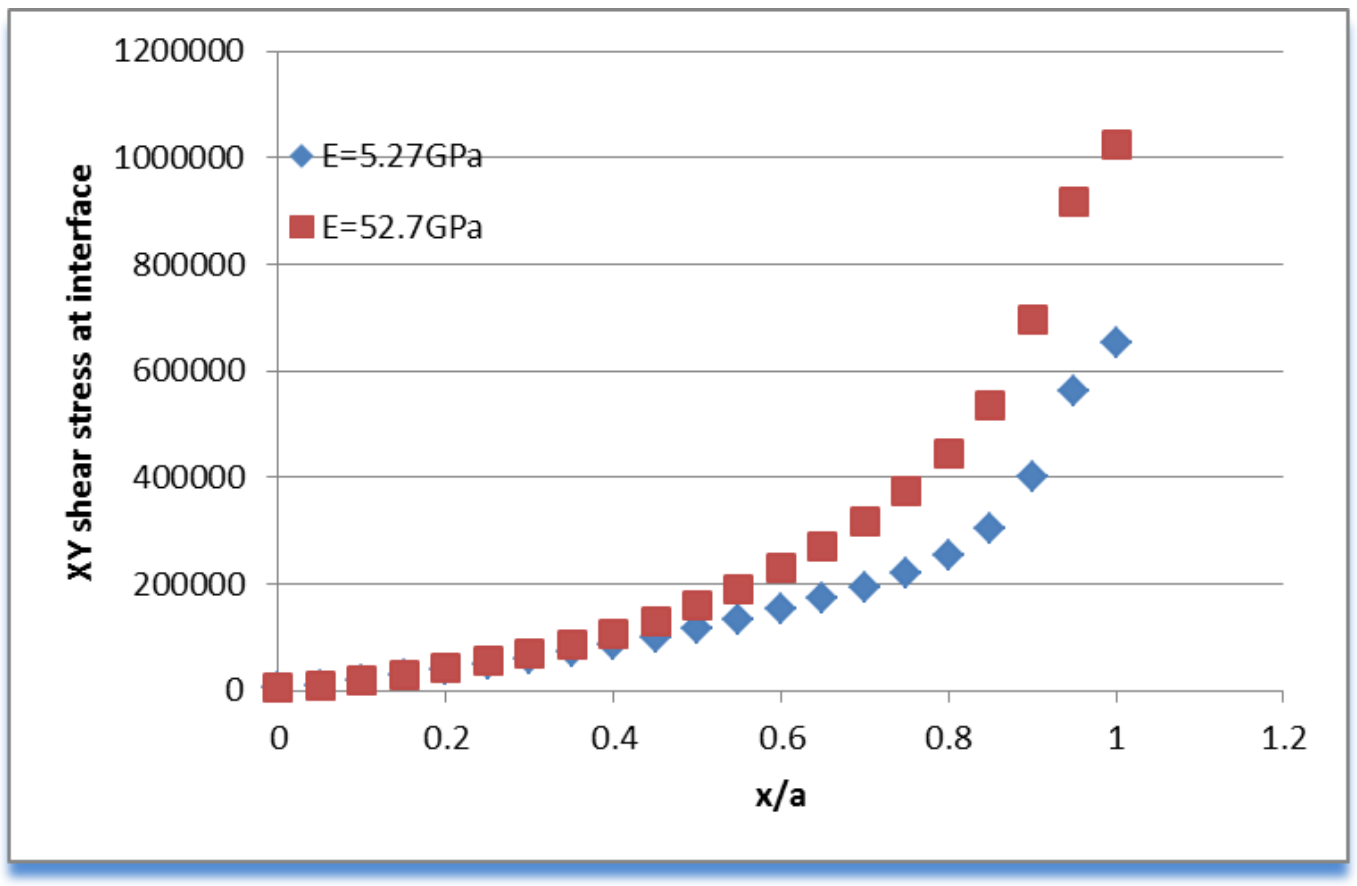

Fig.18 Effect of material properties

It is clearly that the Young's modulus of structure has a significant influence on XY shear stress transferred to host structure. The increase of Young's modulus will result in improving the actuator effect.

This project studies only the effect of Young's modulus of structure as an example to show the method. In terms of other material properties, the model and method are the same.

### *3.2.5 Effect of the bonding layer*

In the previous models, however, the actuator is assumed to be perfectly bonded to the host structures. Typically, piezoelectric sensors are bonded to the host structure by epoxy or conductive epoxy. As a result, a bonding layer will be generated. Since the modulus of the bonding layer is usually lower than that of the sensors and the host structure, it may significantly affect the local stress distribution. In this part, the effect of bonding layer will be studied.

A bonding layer of Epoxy E51-618 is involved, as the light blue part in the fig.19. The shear modulus and Poisson's ratio are 1GPa and 0.38, respectively, and the thickness is 0.25mm. After applying the same boundary conditions and voltage, the results and solutions are shown in fig.20-23

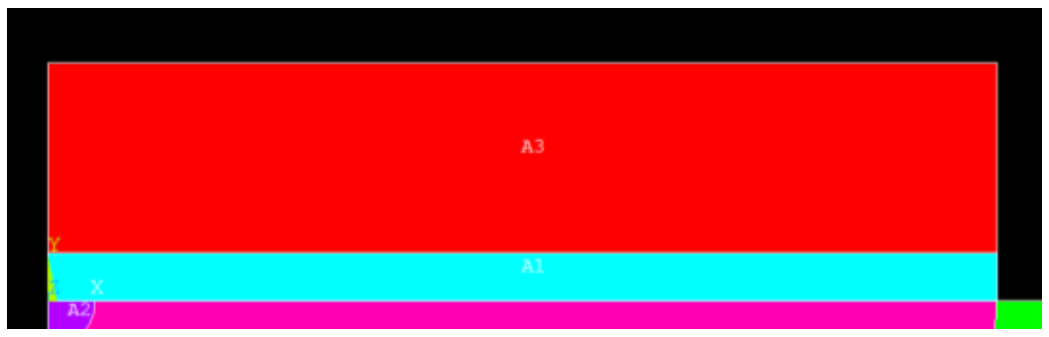

Fig.19 bonding layer geometry (light blue part is bonding layer)

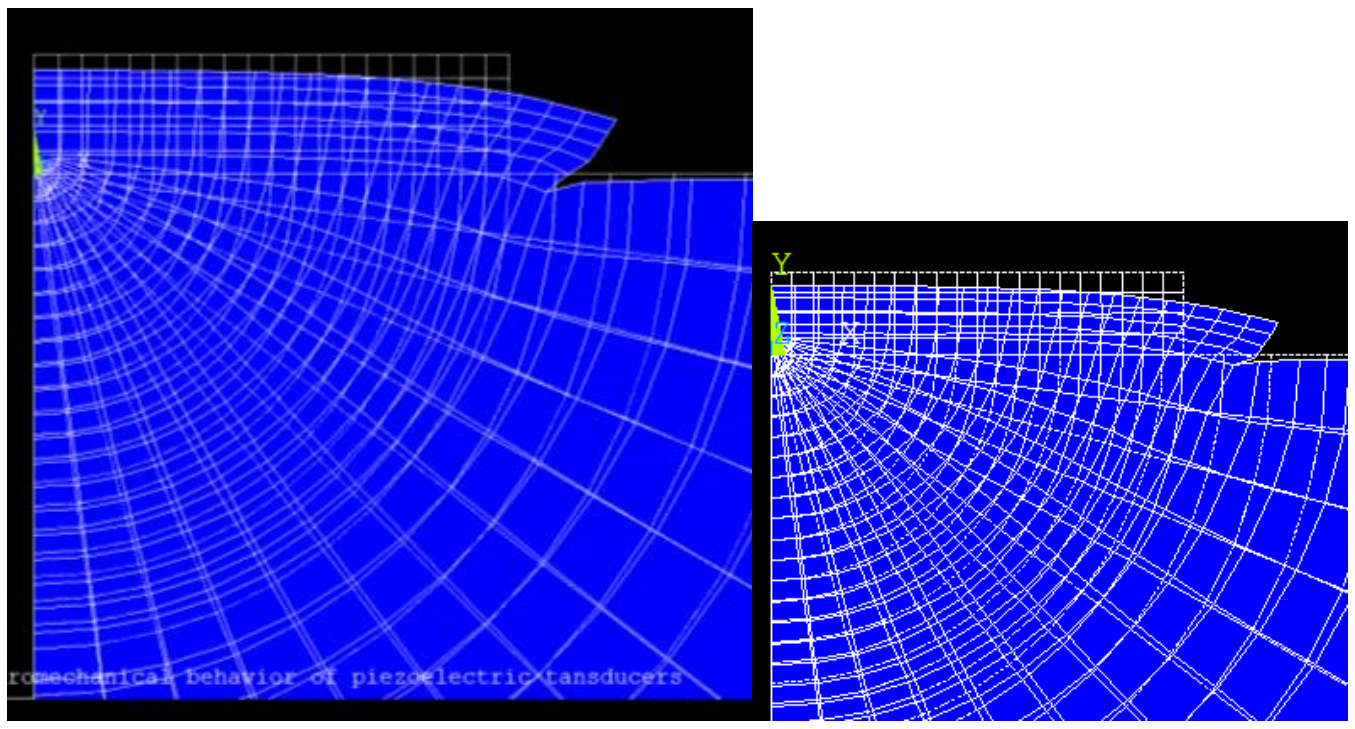

Fig.20 Deformed shape (left: involving bonding layer, right: basic model)

Corresponding author: huangcha@ualberta.ca, Department of Mechanic Engineering, University of Alberta, Canada

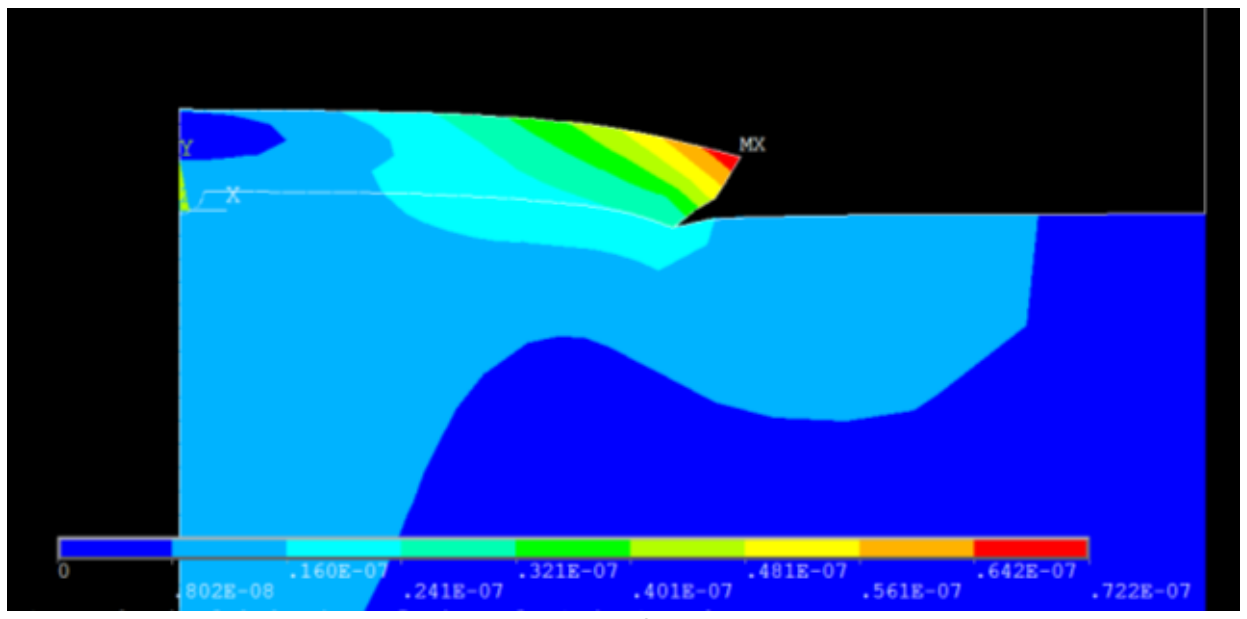

Fig.21 Vector displacement after involving bonding layer

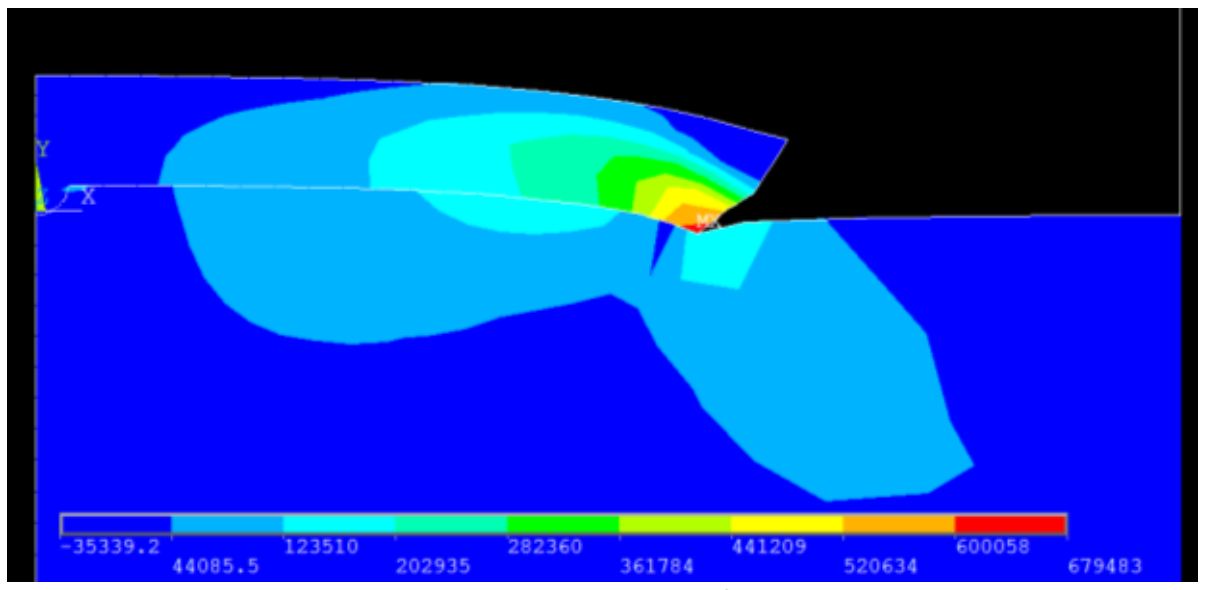

Fig.22 XY shear stress---nodal solution after involving bonding layer

By comparing the deformed shape basic model and model involving the bonding layer in fig .20, there is some decease in deformation after involve bonding layer, but this is not a big difference. Further, the exact shear stresses at interface are list in the following table, and are plotted in fig.23 with the static basic model result.

Table6. Stress result at interface (involving bonding layer)

| x/a | NODE | SX | SY | SXY |
|---|---|---|---|---|
| 0.05 | 876 | 695100 | 20151 | 18845 |
| 0.1 | 879 | 790850 | 42004 | 24957 |
| 0.15 | 880 | 781390 | 32770 | 37216 |
| 0.2 | 881 | 767790 | 37650 | 42962 |
| 0.25 | 882 | 761620 | 39675 | 52926 |
| 0.3 | 883 | 755250 | 43705 | 64017 |
| 0.35 | 884 | 747760 | 48256 | 76002 |
| 0.4 | 885 | 738040 | 53759 | 88722 |
| 0.45 | 886 | 725140 | 59912 | 102120 |
| 0.5 | 887 | 708020 | 66460 | 116141 |
| 0.55 | 888 | 685540 | 72837 | 130723 |
| 0.6 | 889 | 656240 | 78030 | 145903 |
| 0.65 | 890 | 618130 | 80584 | 161957 |
| 0.7 | 891 | 567950 | 77753 | 179805 |
| 0.75 | 892 | 500140 | 67899 | 201241 |
| 0.8 | 893 | 404100 | 43051 | 230350 |
| 0.85 | 894 | 258840 | 17535 | 272237 |
| 0.9 | 895 | -3785 | -74711 | 370000 |
| 0.95 | 896 | -626520 | 13709 | 528000 |

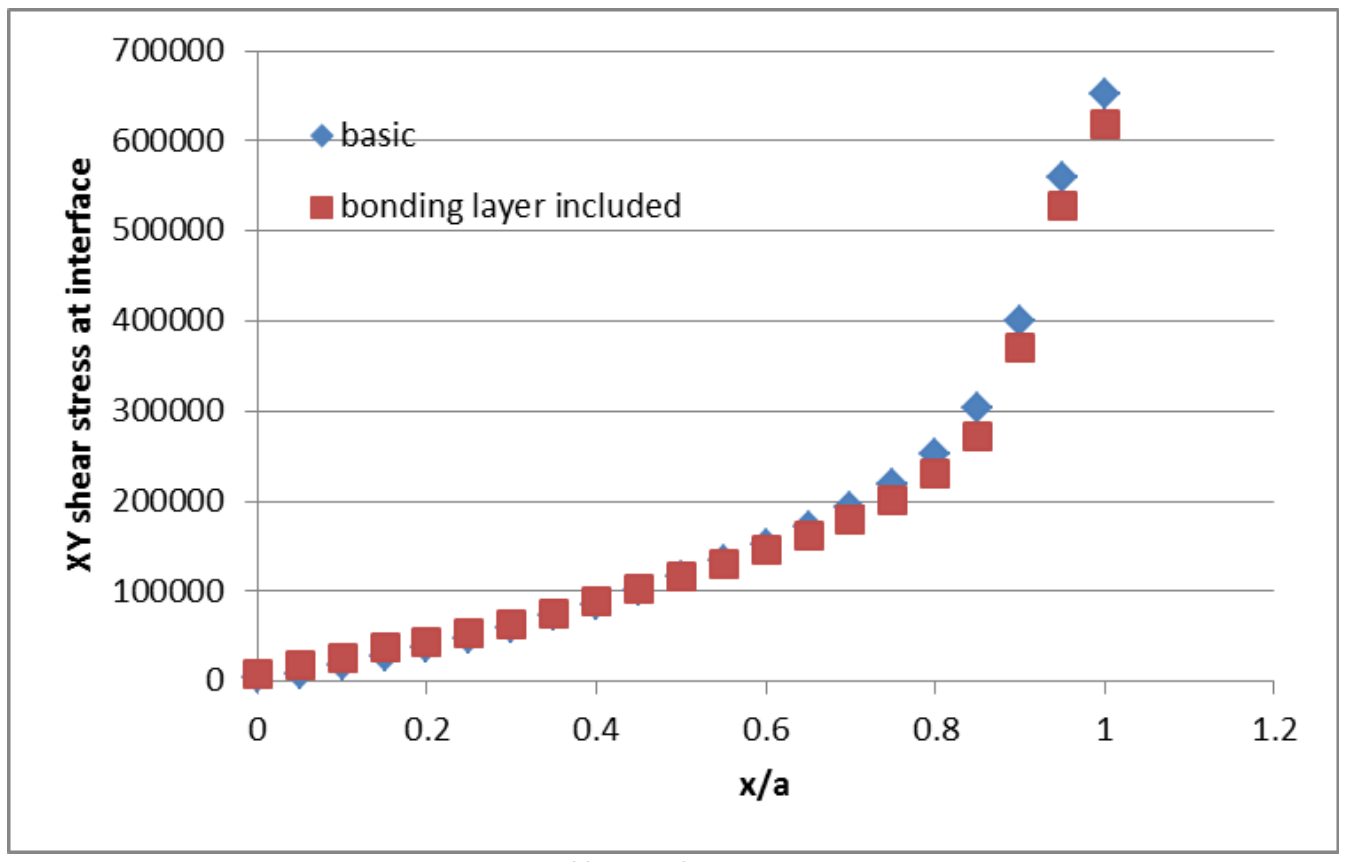

Fig.23 Effect of bonding layer

In fig.20 and 23, the deformation and XY shear stress at interface of bonding layer model are compared with basic model whose actuator is perfectly bonded. As a result, the bonding layer decreases the actuation effect of piezo-actuator but not very clearly in this analysis. After deeper analysis, this is maybe caused by the ignoring of viscoelasticity of bonding layer.

### *3.3 Dynamic analysis and result*

#### *3.3.1 Harmonic analysis*

In the previous studies, however, only static analyses are considered. In this part, the dynamic coupling between the actuator and the host structure will be researched.

The analysis settings are as follows:

Table7. Dynamic analysis settings of ANSYS

| | |
|---|---|
| Analytical type | harmonic |
| Material properties added | density(PZT4) =7500kg/m^3<br>density(STRUCTURE)=7800Kg/m^3 |
| Solution method | full |
| DOF printout format | real + imaginary |
| Load | voltage in upper surface change with frequency |
| Load step options | frequency range: 0-100KHz,<br>number of sub-steps: 100<br>pattern: stepped |

After solution, the stress-frequency relation can be plotted, as shown in Fig.24. The shear stress transferred to the structure changes with the frequency. There are some frequencies near intrinsic mode will cause very large stress.

Corresponding author: huangcha@ualberta.ca, Department of Mechanic Engineering, University of Alberta, Canada

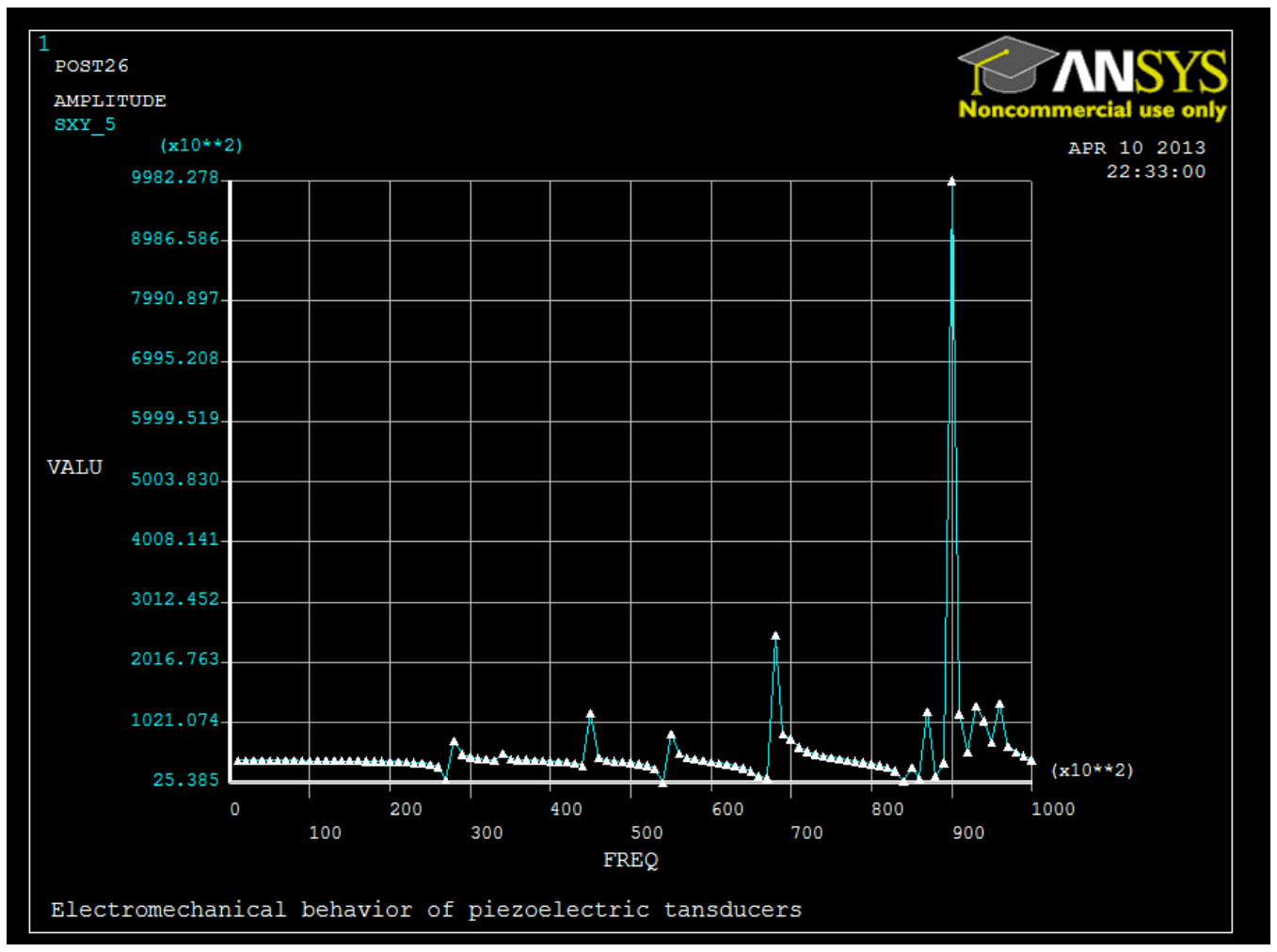

Fig.24 shear stress-frequency of node 180 (near the actuator tip )

### 3.3.2 *Effect of frequency*

Further, the effect of frequency can be studied by comparing the transferred shear stress when load frequency is 100 KHz with static basic model result. The shear stress at interface are listed in table8 and plotted in fig.25. It is clearly that the frequency has a significant effect in the stress transferred from the actuator to the host structure.

Besides, other results of different frequencies and different nodes are list in the appendices, like the shear stress along the -45 degree line.

Table8. XY shear stress at interface

| x/a | Freq.=0 | Freq.=1000000 |
|---|---|---|
| 0.05 | 0 | 1000000 |
| 0.1 | 8892 | 32567.54 |
| 0.15 | 18156 | 65914.57 |
| 0.2 | 27727 | 99736.31 |
| 0.25 | 37705 | 133622.4 |
| 0.3 | 48375 | 167290.3 |
| 0.35 | 59846 | 200246.3 |
| 0.4 | 72132 | 232085.3 |
| 0.45 | 85393 | 262299.7 |
| 0.5 | 1.00E+05 | 290266.1 |
| 0.55 | 1.16E+05 | 315366.4 |
| 0.6 | 1.34E+05 | 345541.6 |
| 0.65 | 1.52E+05 | 372300.6 |
| 0.7 | 1.72E+05 | 393145.5 |
| 0.75 | 1.94E+05 | 420951.4 |
| 0.8 | 2.18E+05 | 446203.1 |
| 0.85 | 2.52E+05 | 480582.2 |
| 0.9 | 3.03E+05 | 521012.9 |
| 0.95 | 4.00E+05 | 612972 |

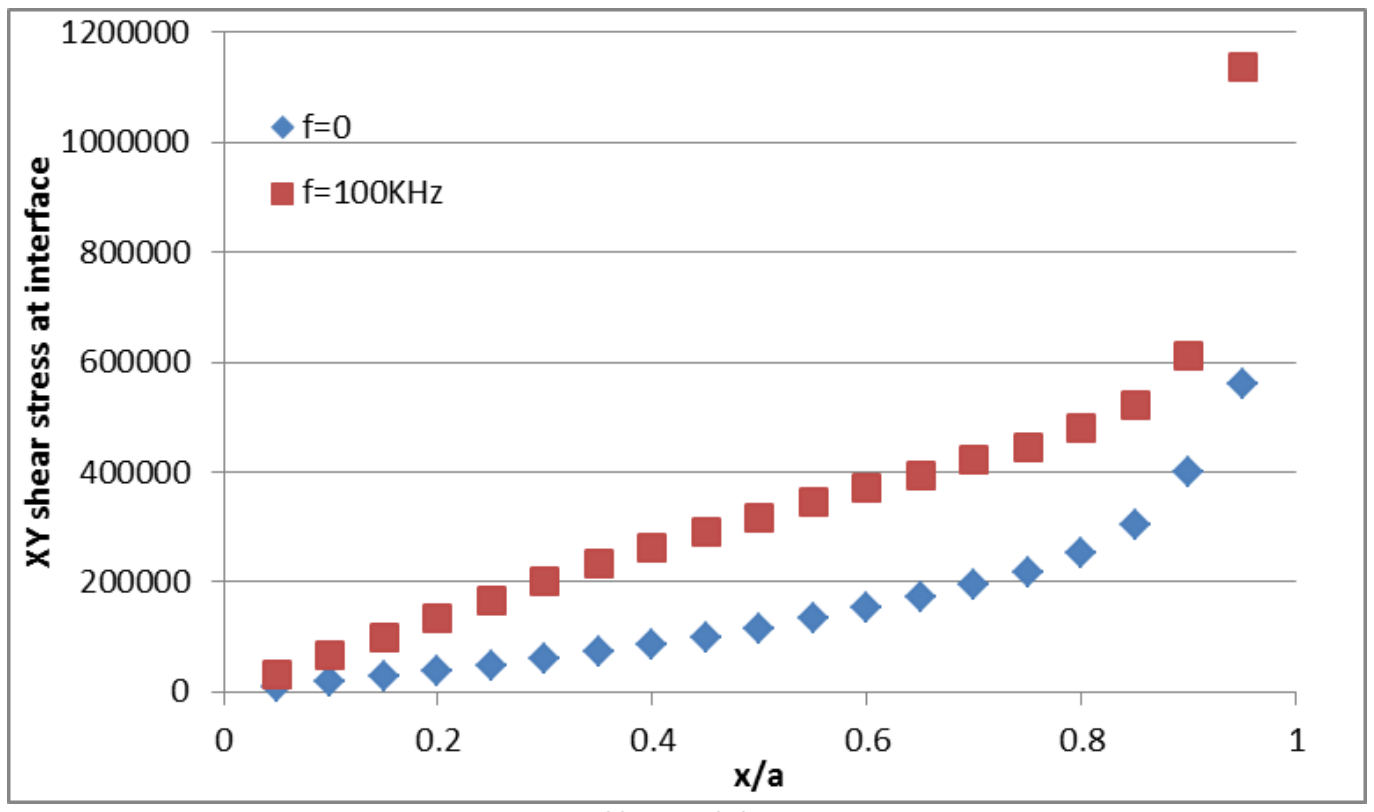

Fig.25 effect of frequency

## 4 CONCLUSIONS

After previous analysis and discussion, following conclusions can be summarized:

1) Stress and displacement transferred from the actuator are determined and there is a singularity at the tips of the actuator.
2) Material properties and frequency have a significant influence on the stresses transferred to the host structure, for example, the increase of Young's modulus will result in improving the actuator effect.
3) Bonding layer decreases the stress transferred to the structure.
4) FEM is good method to study the effect of material properties, bonding layer and frequency in piezoelectric actuator models.

Corresponding author: huangcha@ualberta.ca, Department of Mechanic Engineering, University of Alberta, Canada

Corresponding author: huangcha@ualberta.ca, Department of Mechanic Engineering, University of Alberta, Canada

## APPENDICES

### 1. Basic model results List

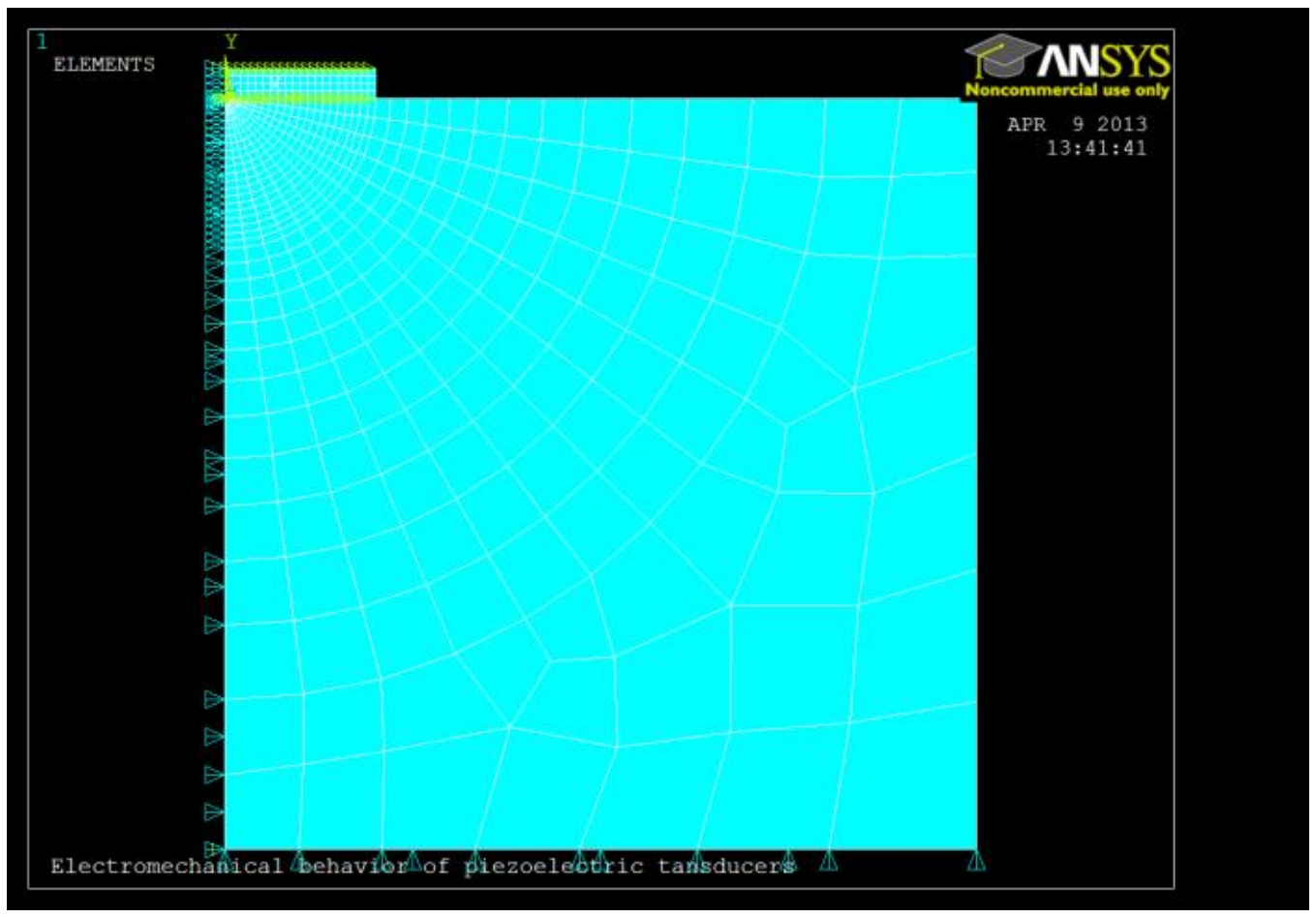

Figure A1. Mesh result and boundary conditions (basic model)

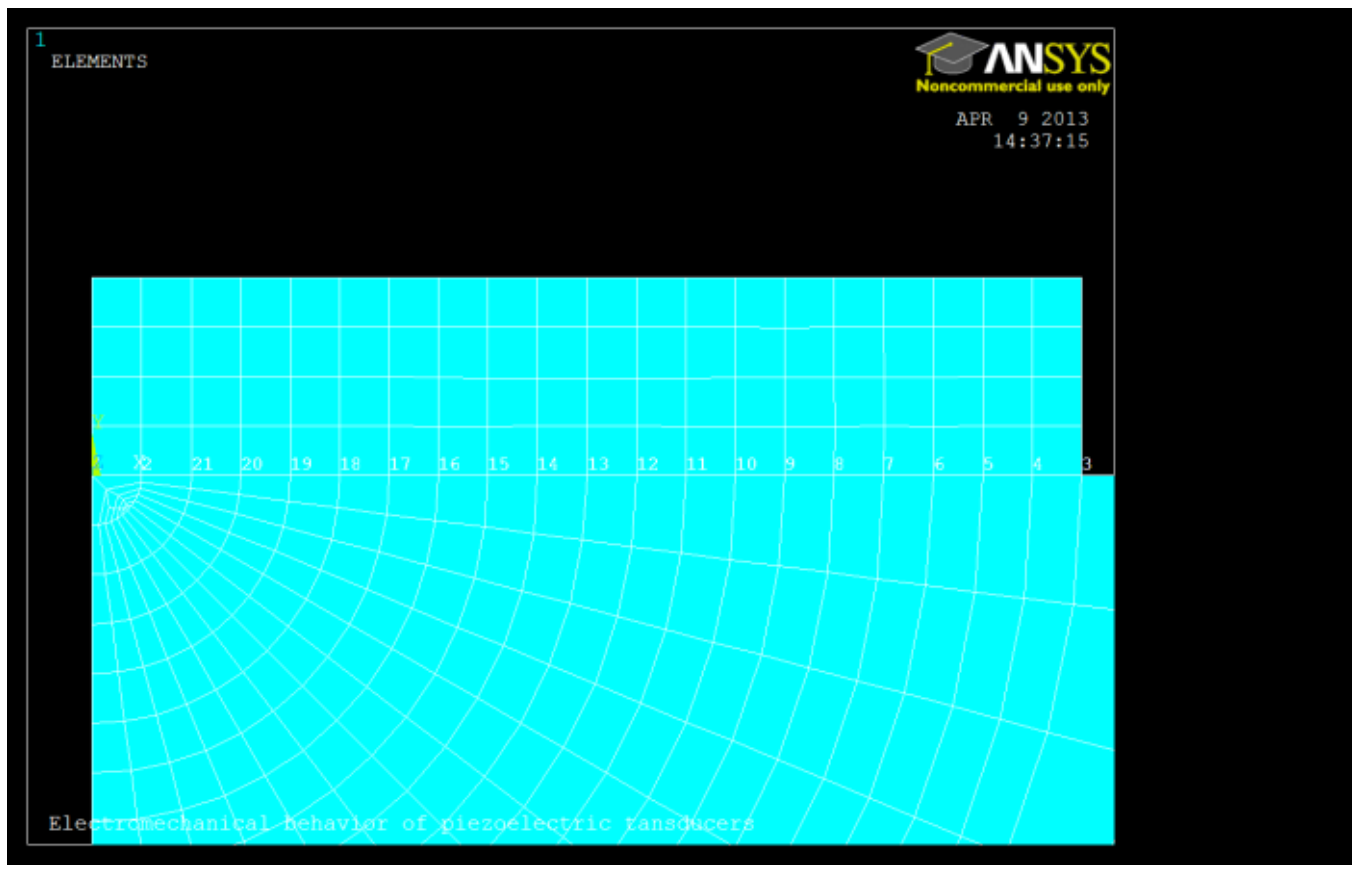

Figure A2. Interface nodal number (basic model)

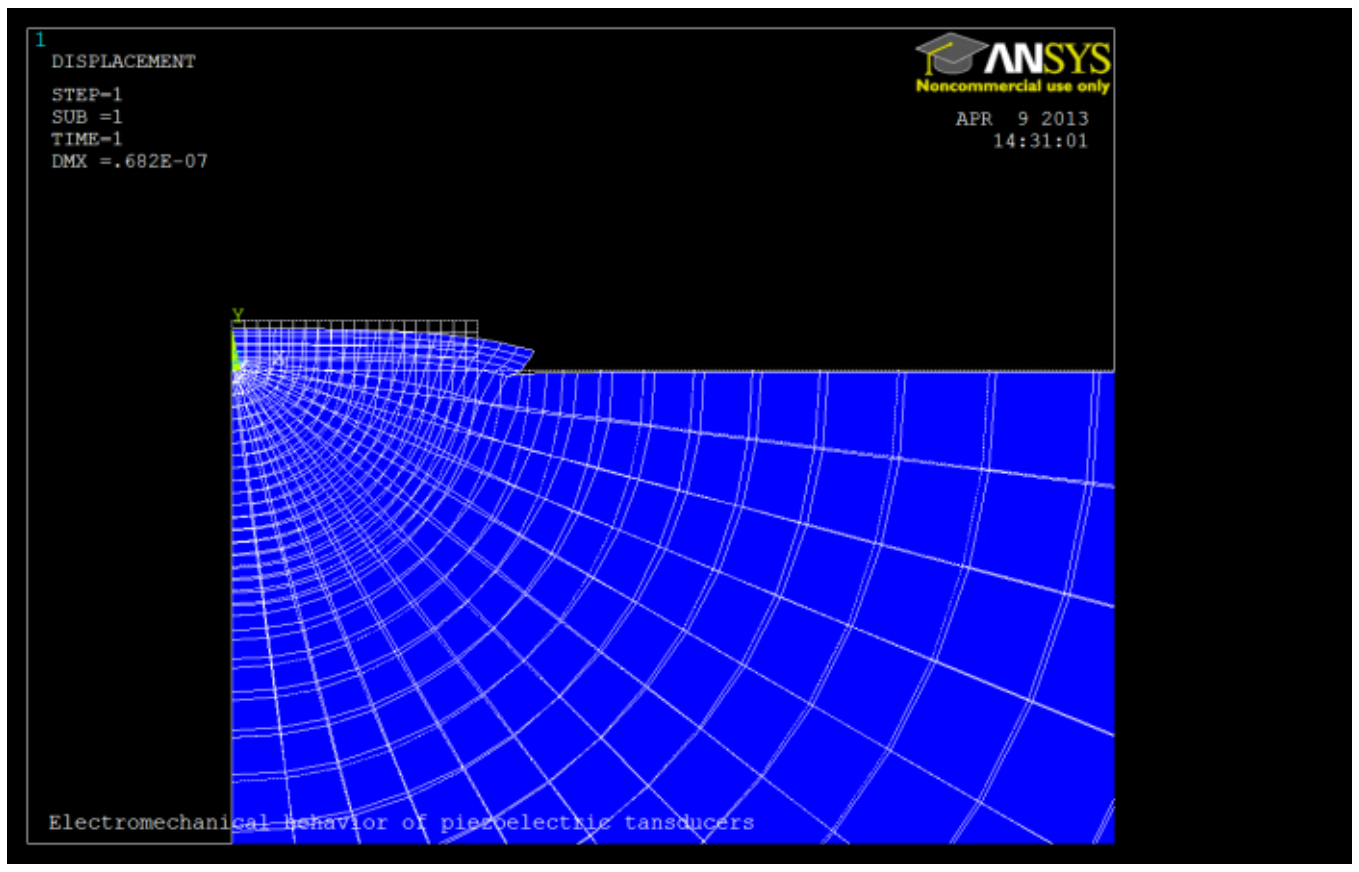

Figure A3. Deformed shape (basic model)

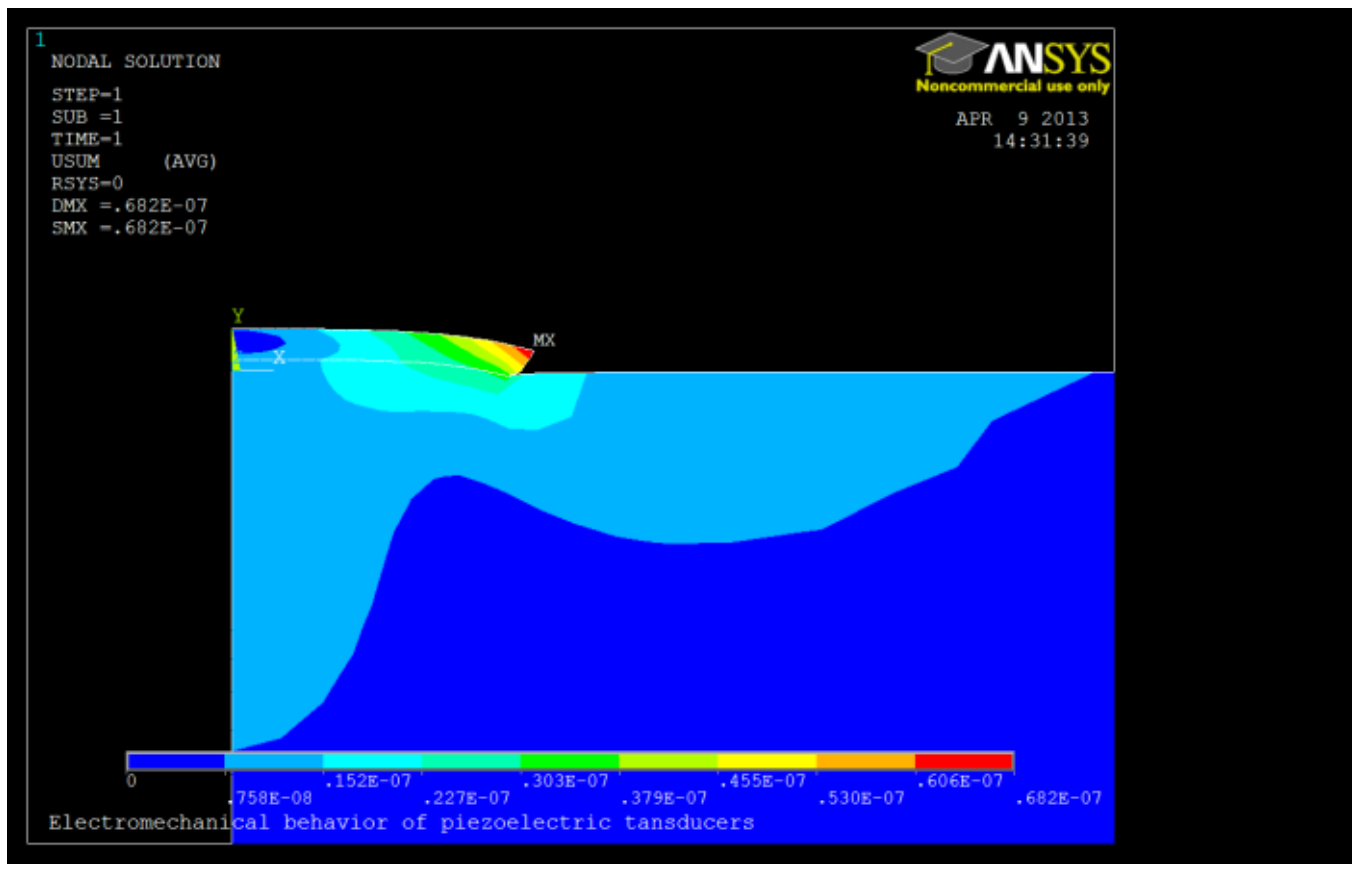

Figure A4. Vector displacement ---nodal solution (basic model)

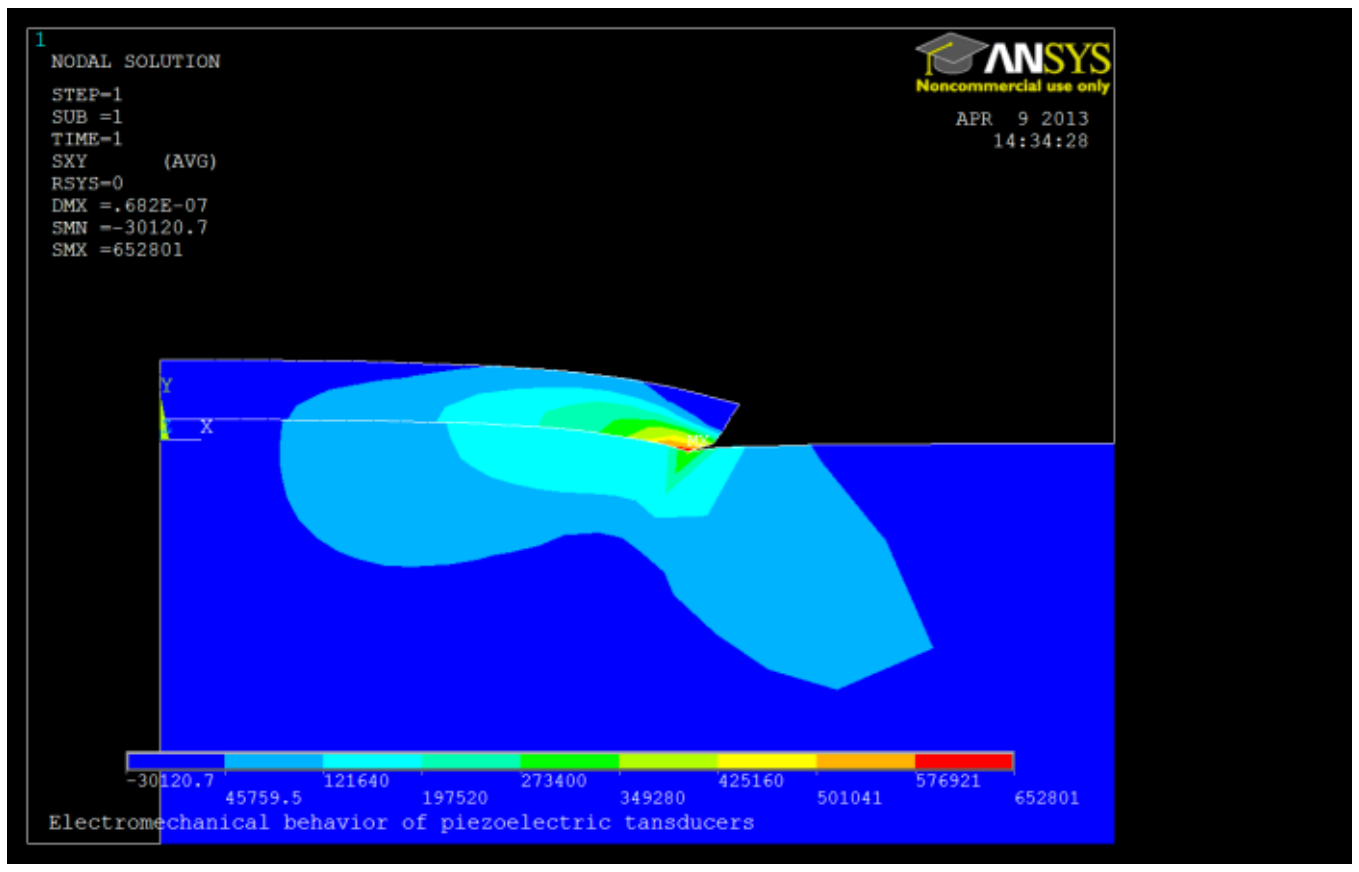

Figure A5. XY shear stress---nodal solution (basic model)

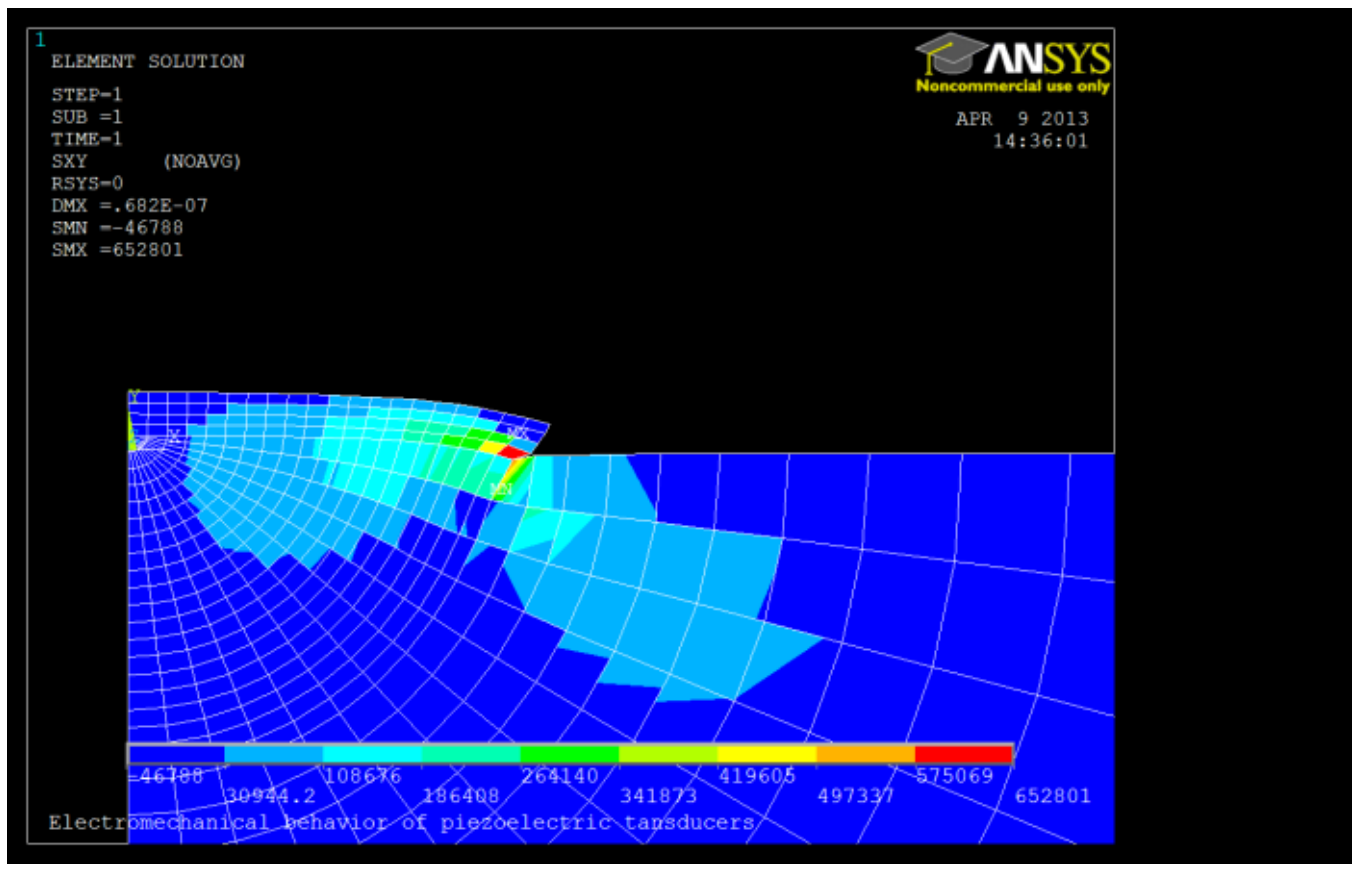

Figure A6. XY shear stress---element solution (basic model)

Corresponding author: huangcha@ualberta.ca, Department of Mechanic Engineering, University of Alberta, Canada

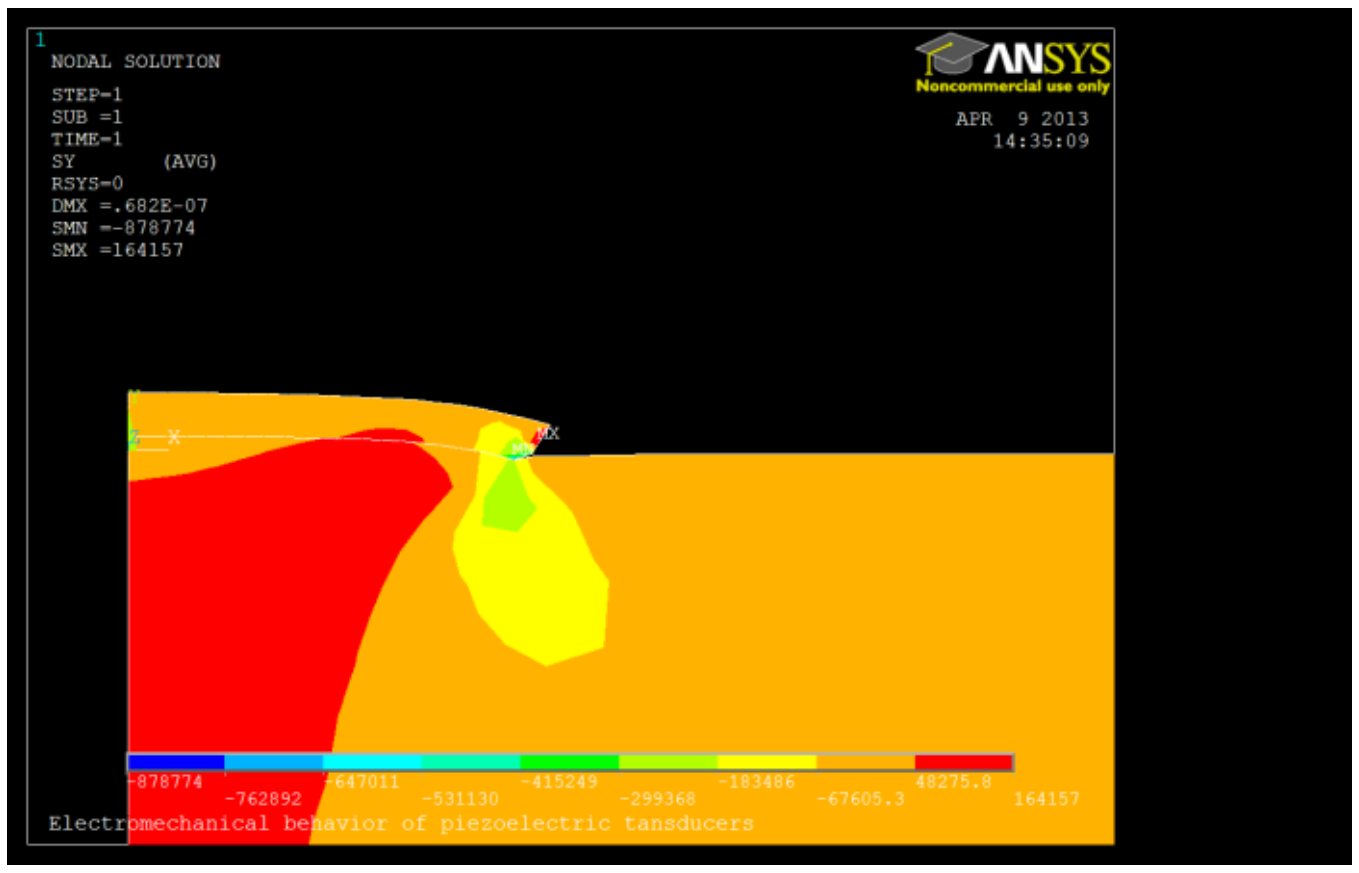

Figure A7. Y stress---nodal solution (basic model)

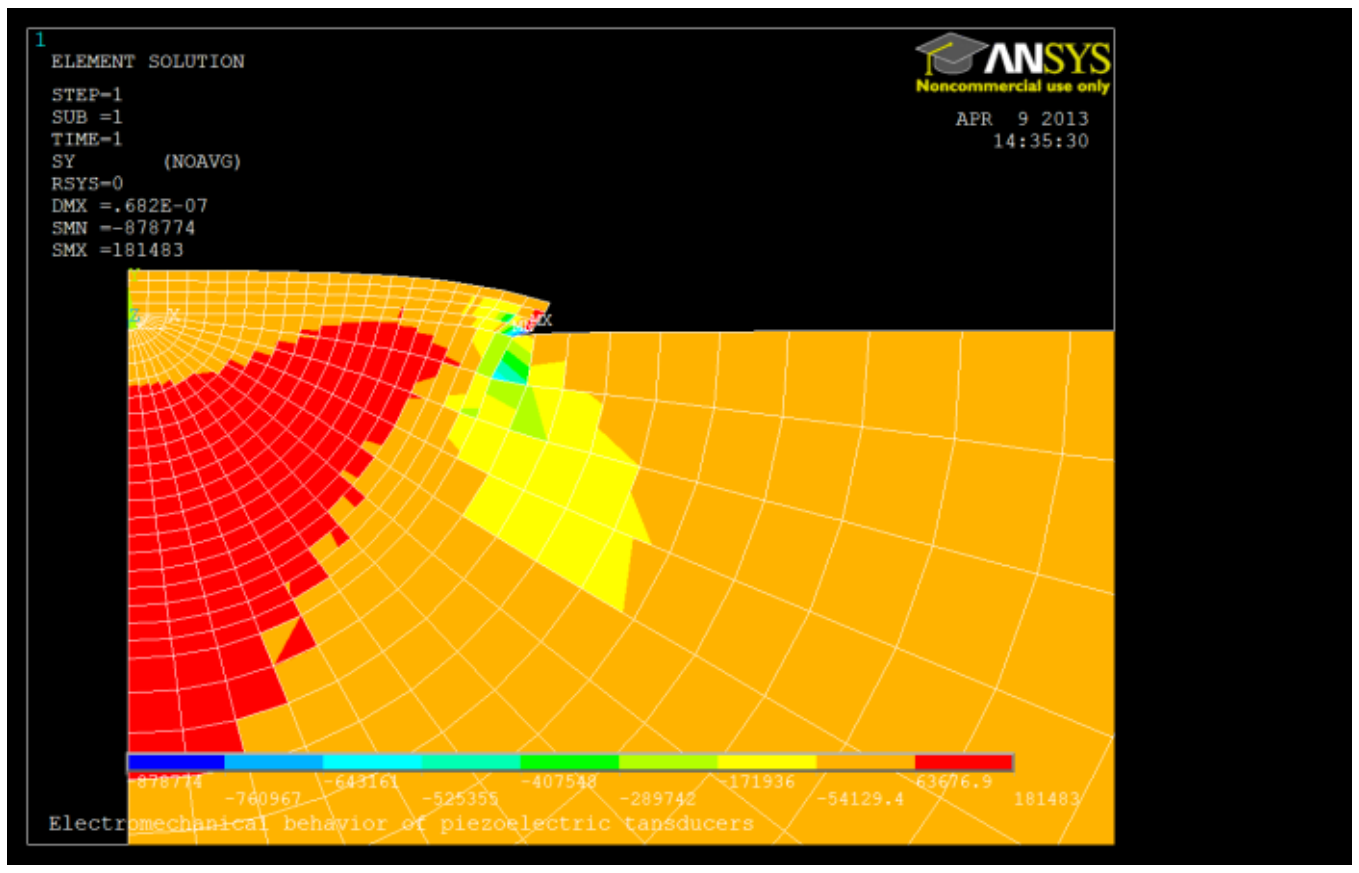

Figure A8. Y stress ---element solution (basic model)

Table A1. Stress result at interface (basic model)

| x/a | NODE | SX | SY | SXY |
|---|---|---|---|---|
| 0.05 | 2 | -1021900 | 19046 | 8892 |
| 0.1 | 21 | -1021500 | 19698 | 18156 |
| 0.15 | 20 | -1020000 | 20781 | 27727 |
| 0.2 | 19 | -1018300 | 22366 | 37705 |
| 0.25 | 18 | -1016400 | 24438 | 48375 |
| 0.3 | 17 | -1015000 | 27267 | 59846 |
| 0.35 | 16 | -1014200 | 30749 | 72132 |
| 0.4 | 15 | -1014800 | 34894 | 85393 |
| 0.45 | 14 | -1017300 | 39478 | 100050 |
| 0.5 | 13 | -1023300 | 45012 | 116160 |
| 0.55 | 12 | -1033300 | 50715 | 133670 |
| 0.6 | 11 | -1049500 | 56493 | 152410 |
| 0.65 | 10 | -1072500 | 60621 | 172320 |
| 0.7 | 9 | -1105100 | 61230 | 193760 |
| 0.75 | 8 | -1149300 | 54679 | 218460 |
| 0.8 | 7 | -1215000 | 32480 | 252300 |
| 0.85 | 6 | -1327400 | 3326.5 | 302770 |
| 0.9 | 5 | -1558800 | -91799 | 400370 |
| 0.95 | 4 | -2101400 | -79717 | 560350 |

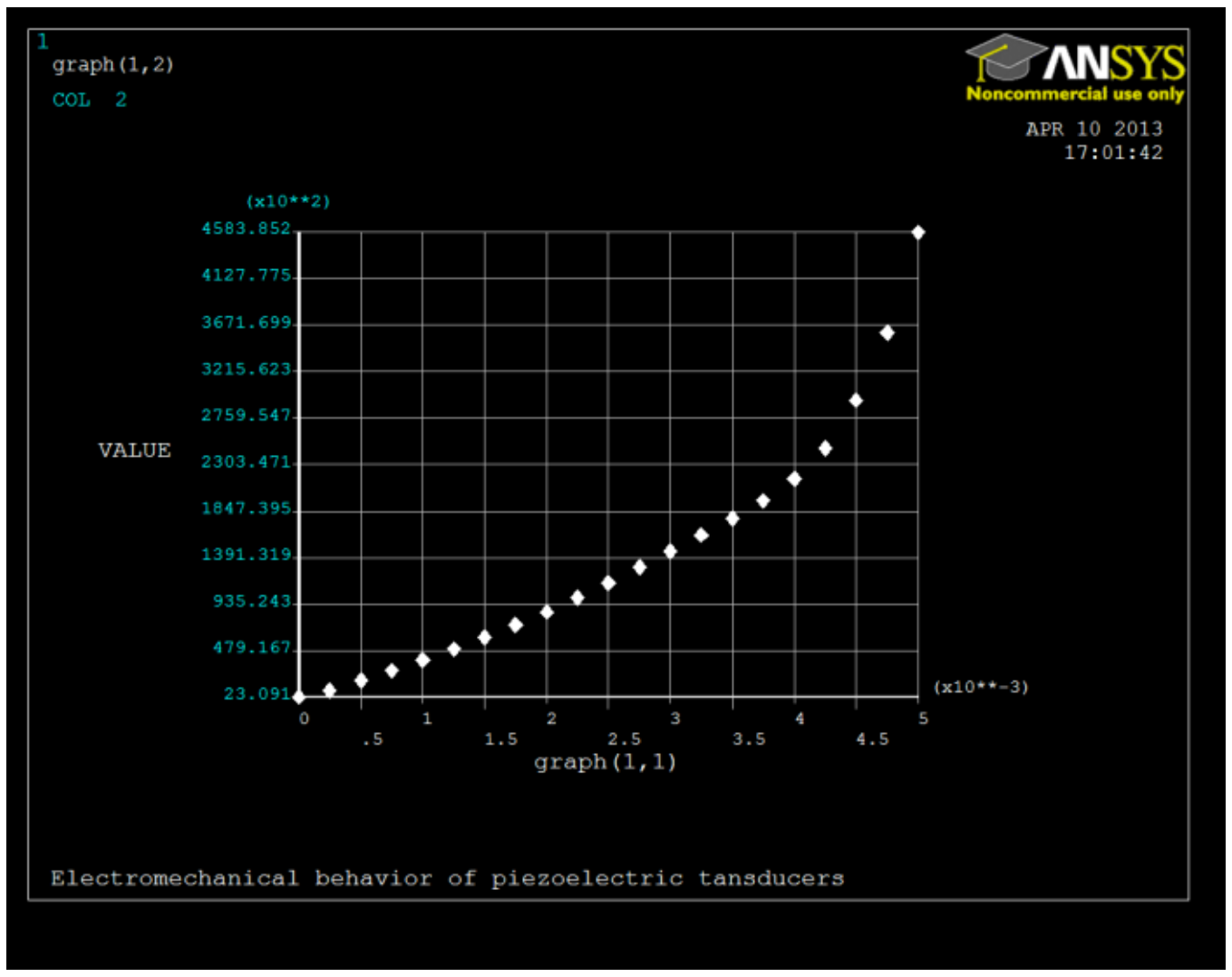

Figure A9. XY shear stress at interface (basic model)

## 2. Simulation result of different material properties (10E)

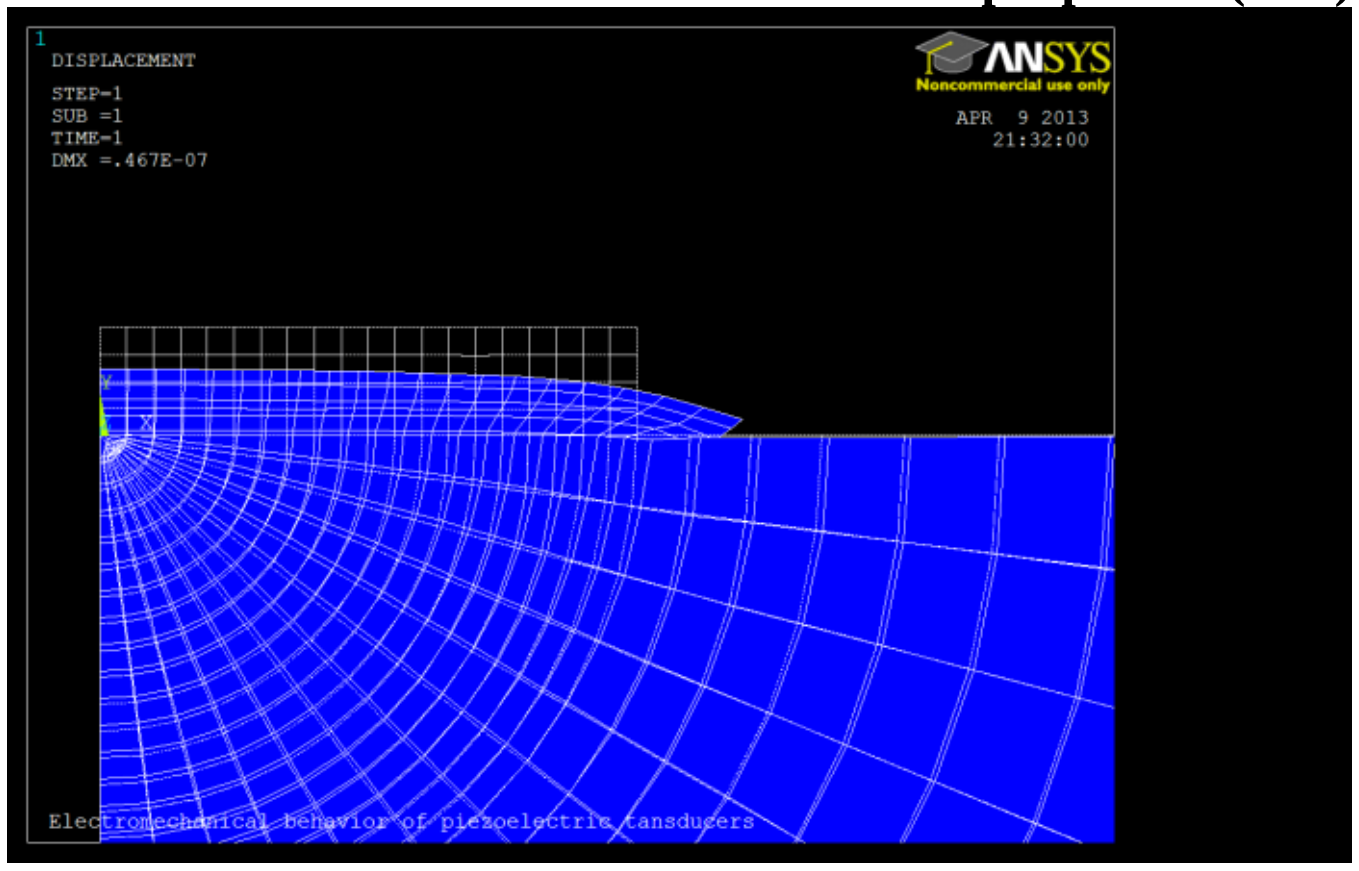

Figure A10. deformed shape (different material properties model)

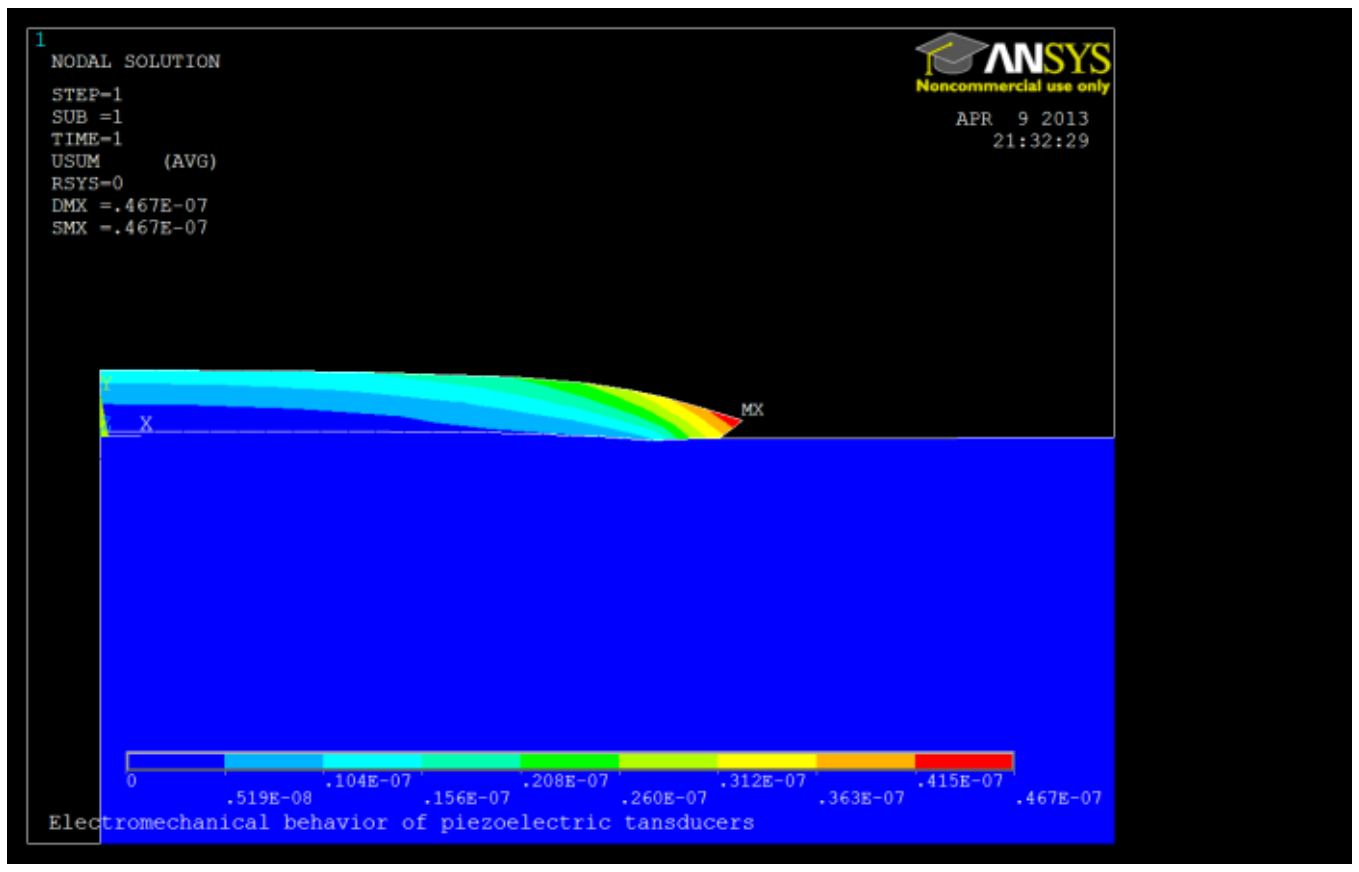

figure A11. Vector displacement (different material properties model)

Corresponding author: huangcha@ualberta.ca, Department of Mechanic Engineering, University of Alberta, Canada

APPENDICES

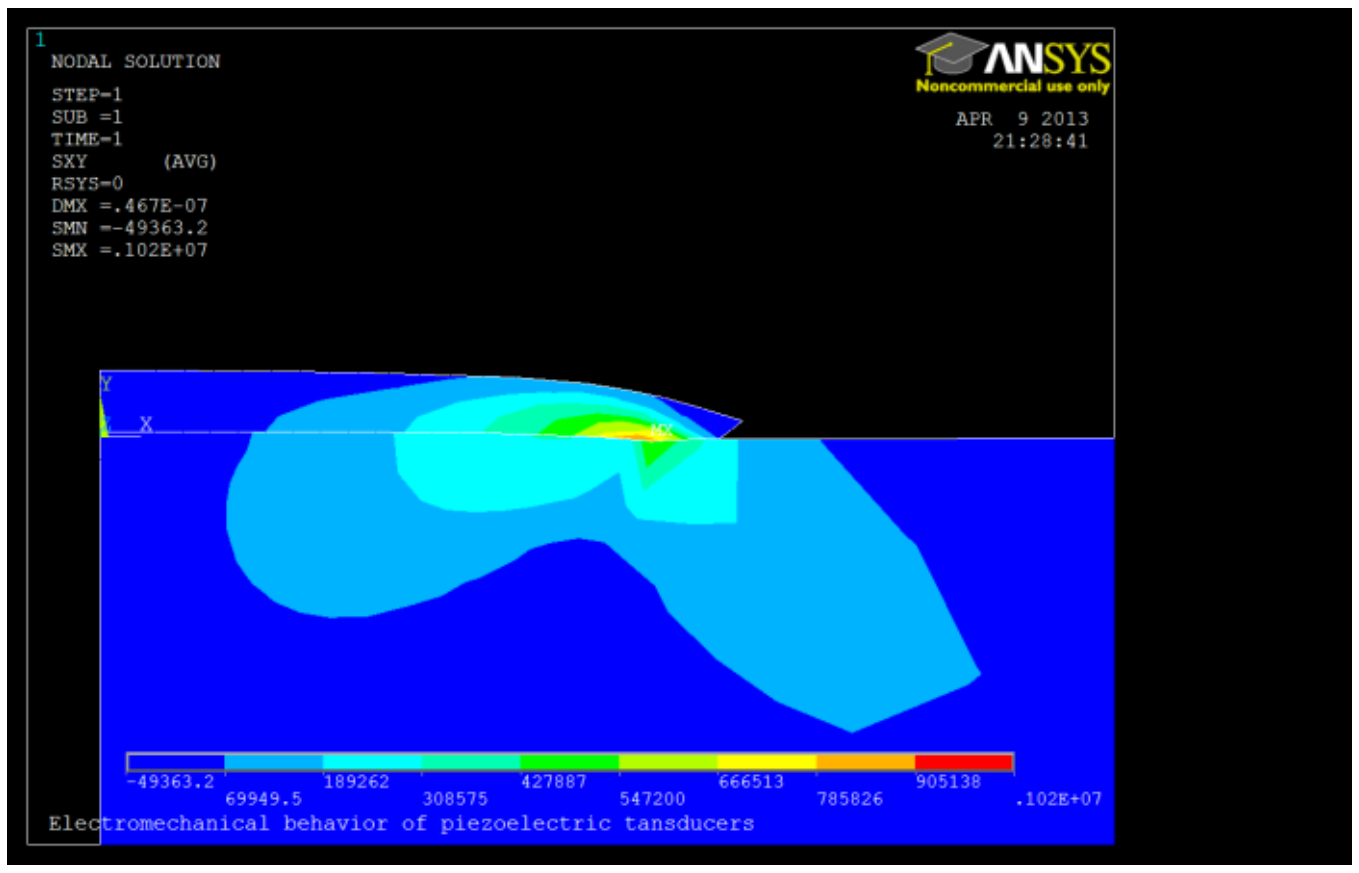

Figure A12. XY shear stress---nodal solution (different material properties model)

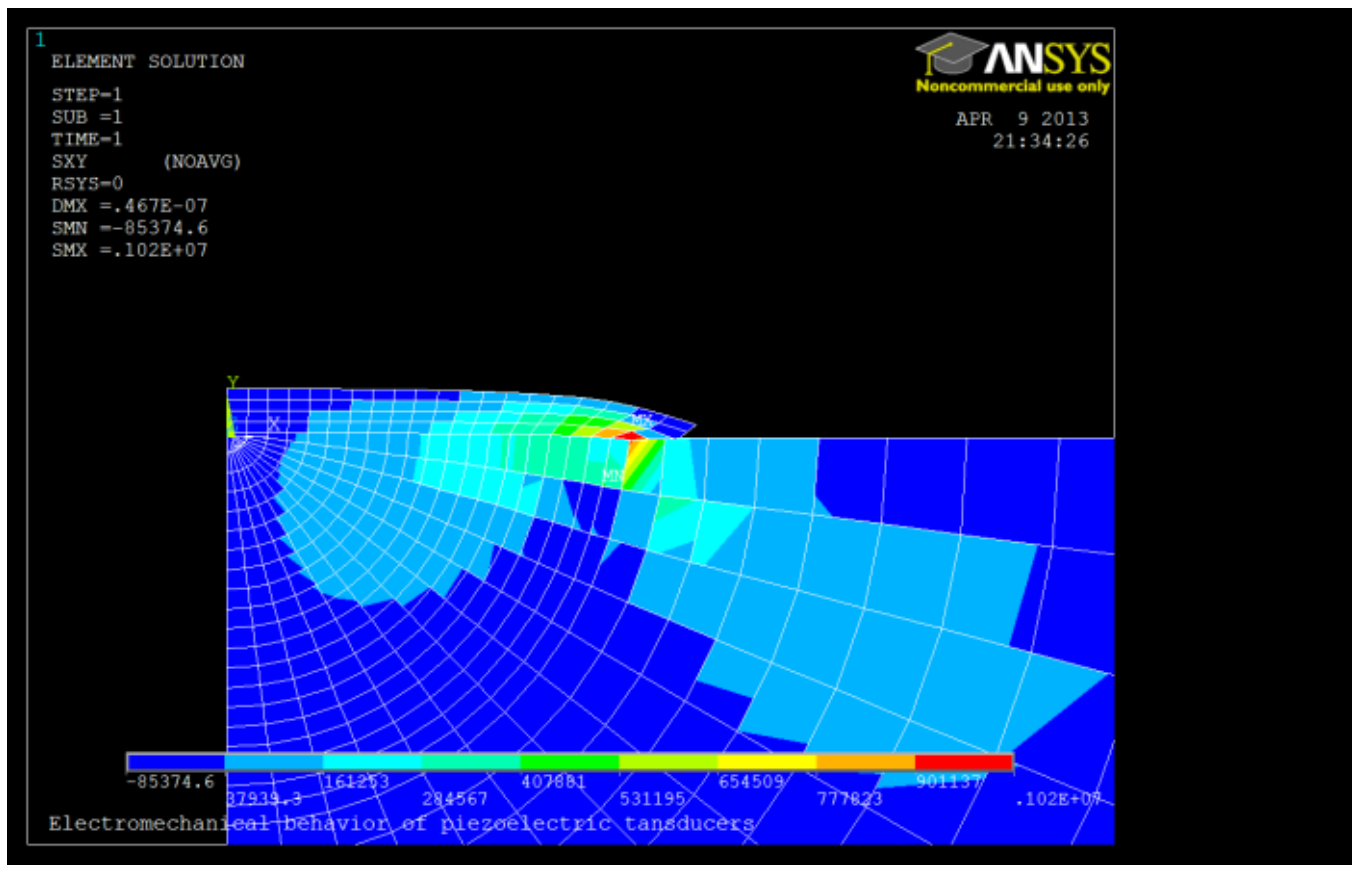

Figure A13. XY shear stress---element solution (different material properties model)

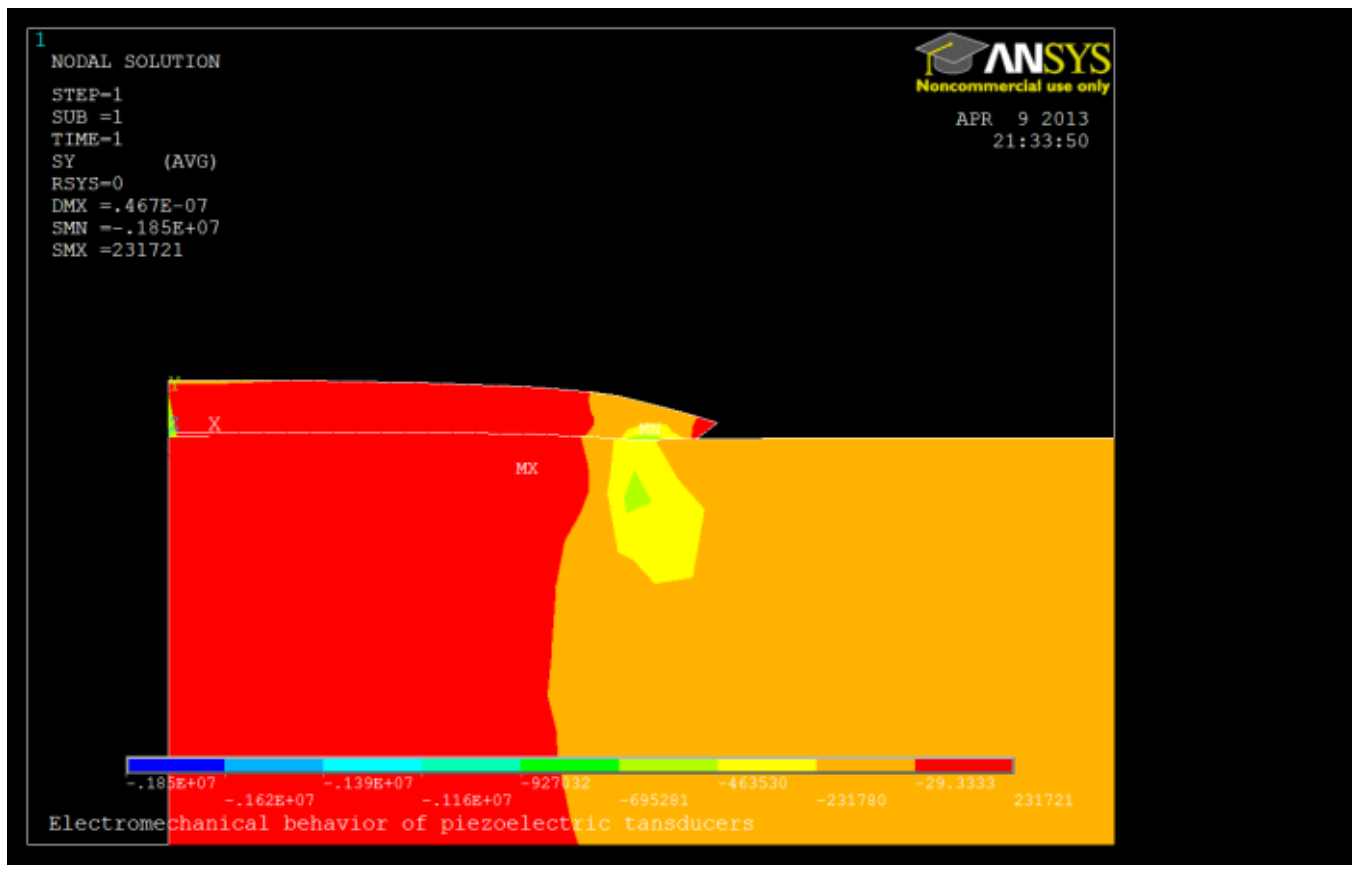

Figure A14. Y stress---nodal solution (different material properties model)

Table A2. Stress result at interface (different material properties model)

| | NODE | SX | SY | SXY |
|---|---|---|---|---|
| **0.05** | 2 | -1.55E+06 | 19892 | 9584.5 |
| **0.1** | 21 | -1.55E+06 | 20871 | 19559 |
| **0.15** | 20 | -1.55E+06 | 22596 | 30187 |
| **0.2** | 19 | -1.55E+06 | 25088 | 41802 |
| **0.25** | 18 | -1.55E+06 | 28488 | 54911 |
| **0.3** | 17 | -1.54E+06 | 33168 | 69909 |
| **0.35** | 16 | -1.54E+06 | 39249 | 87133 |
| **0.4** | 15 | -1.54E+06 | 47017 | 1.07E+05 |
| **0.45** | 14 | -1.53E+06 | 56748 | 1.31E+05 |
| **0.5** | 13 | -1.53E+06 | 69480 | 1.59E+05 |
| **0.55** | 12 | -1.52E+06 | 85074 | 1.91E+05 |
| **0.6** | 11 | -1.52E+06 | 1.04E+05 | 2.28E+05 |
| **0.65** | 10 | -1.52E+06 | 1.24E+05 | 2.71E+05 |
| **0.7** | 9 | -1.54E+06 | 1.41E+05 | 3.19E+05 |
| **0.75** | 8 | -1.56E+06 | 1.48E+05 | 3.74E+05 |
| **0.8** | 7 | -1.62E+06 | 1.31E+05 | 4.44E+05 |
| **0.85** | 6 | -1.75E+06 | 84877 | 5.36E+05 |
| **0.9** | 5 | -1.98E+06 | -56292 | 6.97E+05 |
| **0.95** | 4 | -2.64E+06 | -2.04E+05 | 9.16E+05 |

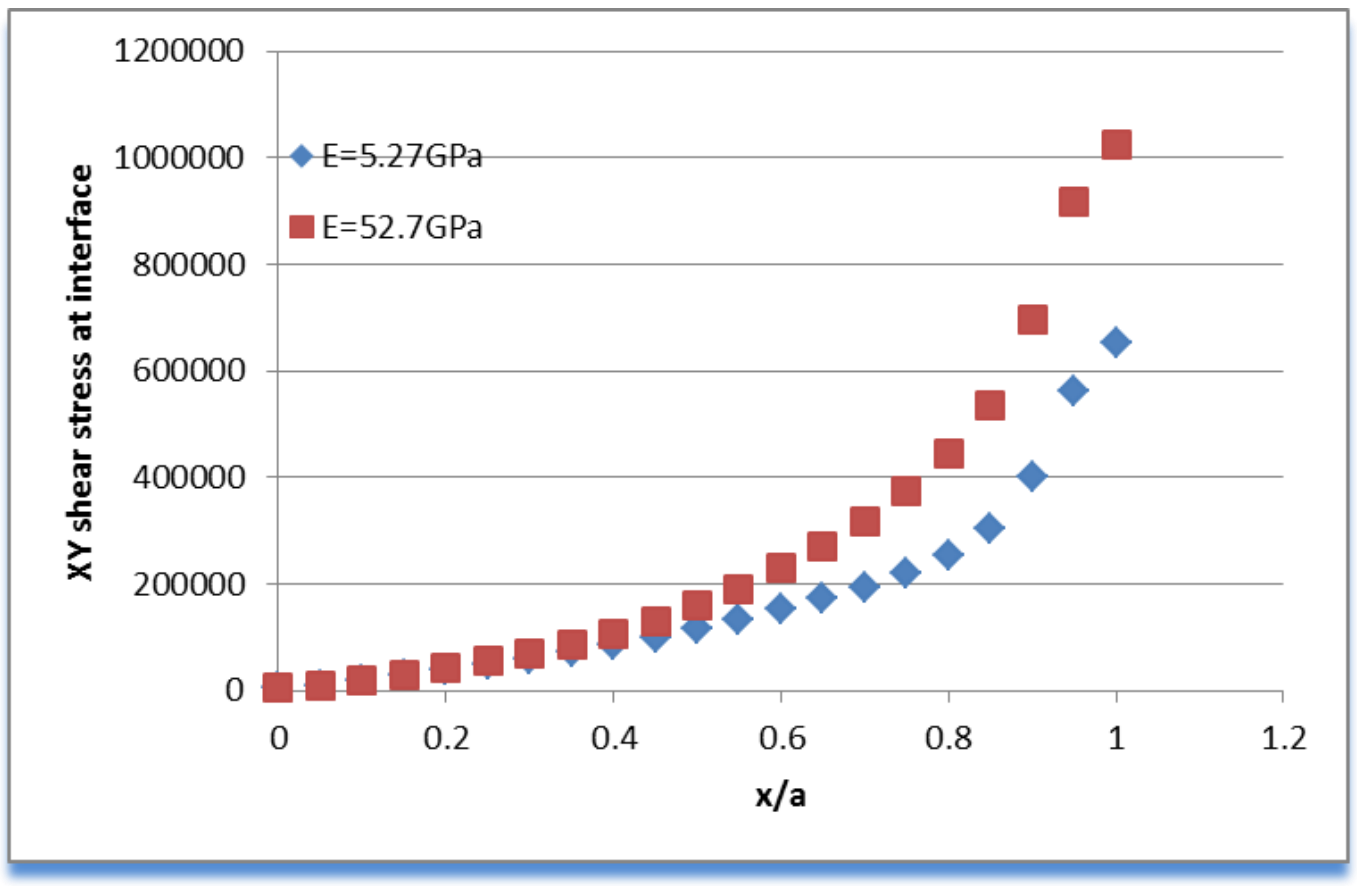

Figure A15. Effect of material properties

### 3. Simulation result involving bonding layer

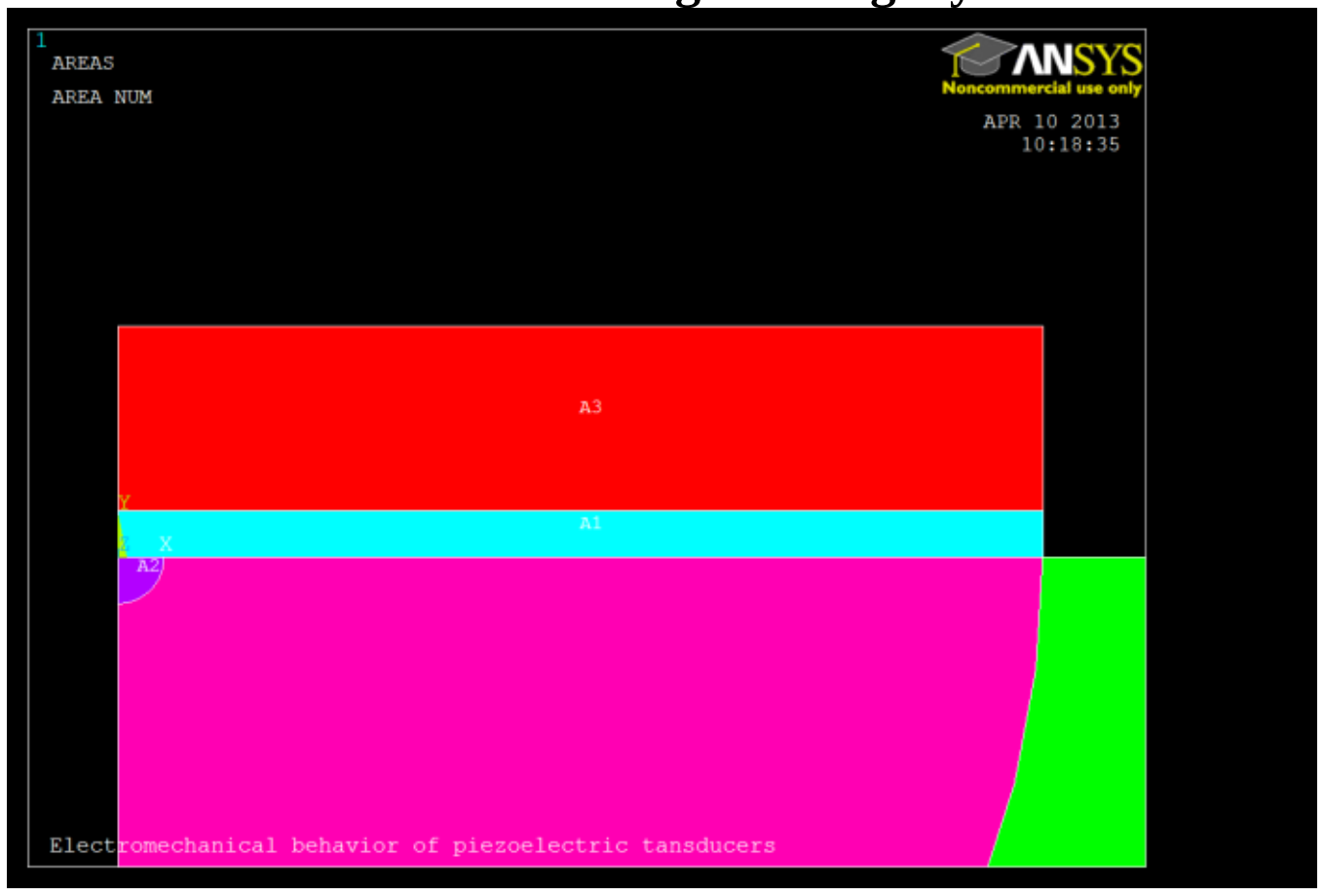

Figure A16. Bonding layer geometry

Corresponding author: huangcha@ualberta.ca, Department of Mechanic Engineering, University of Alberta, Canada

APPENDICES

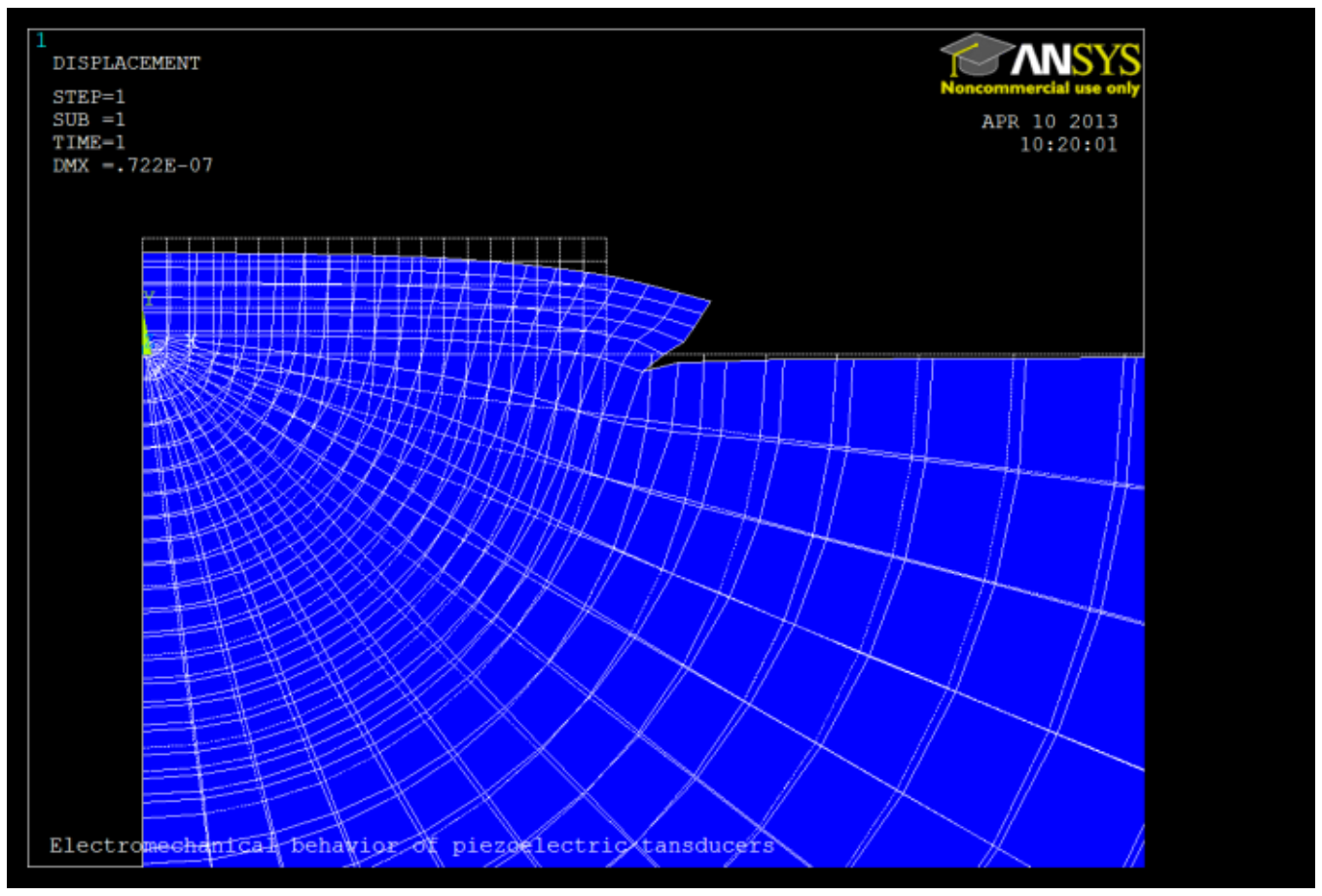

Figure A17. Deformed shape (involving bonding layer)

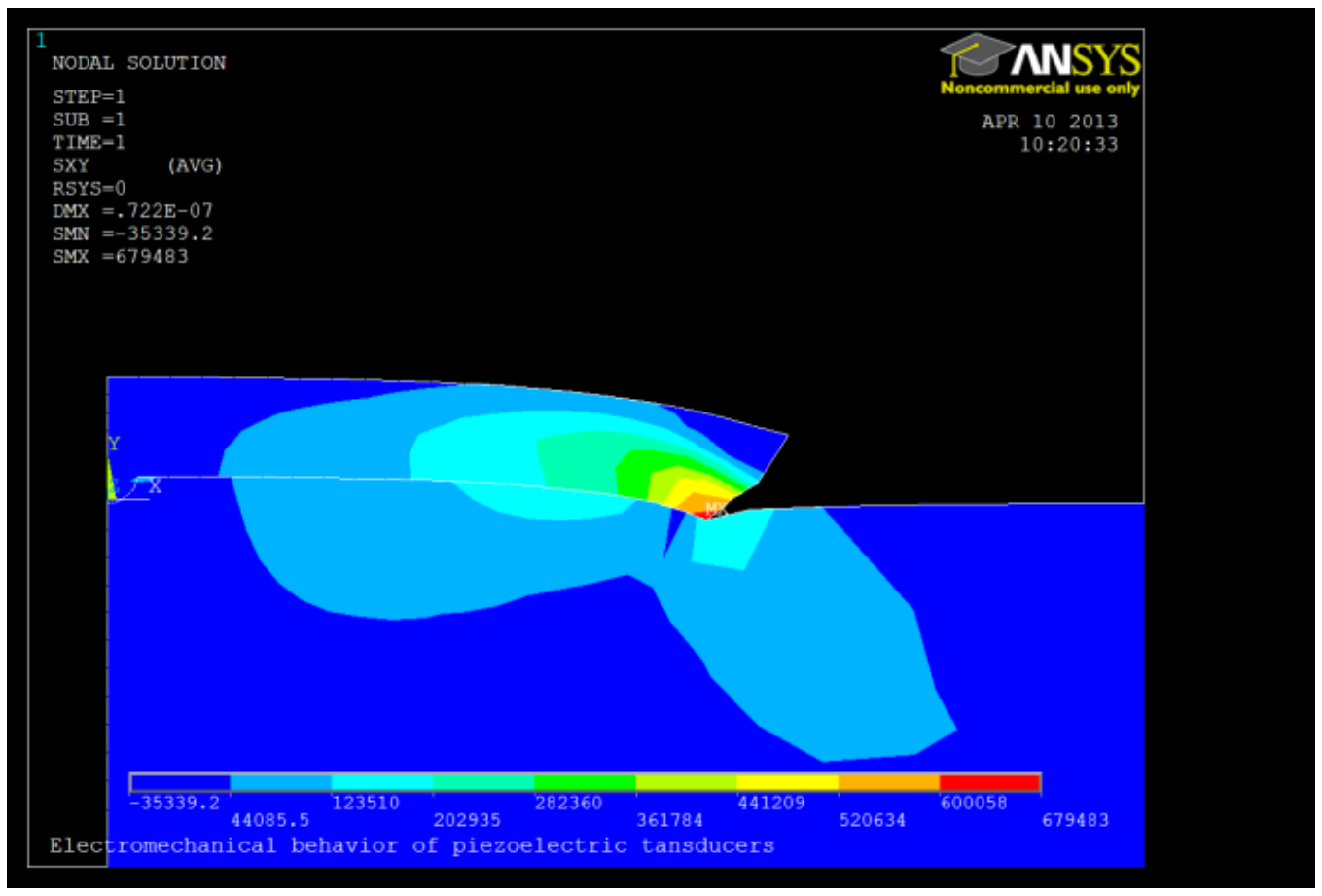

Figure A20. XY shear stress---nodal solution (involving bonding layer)

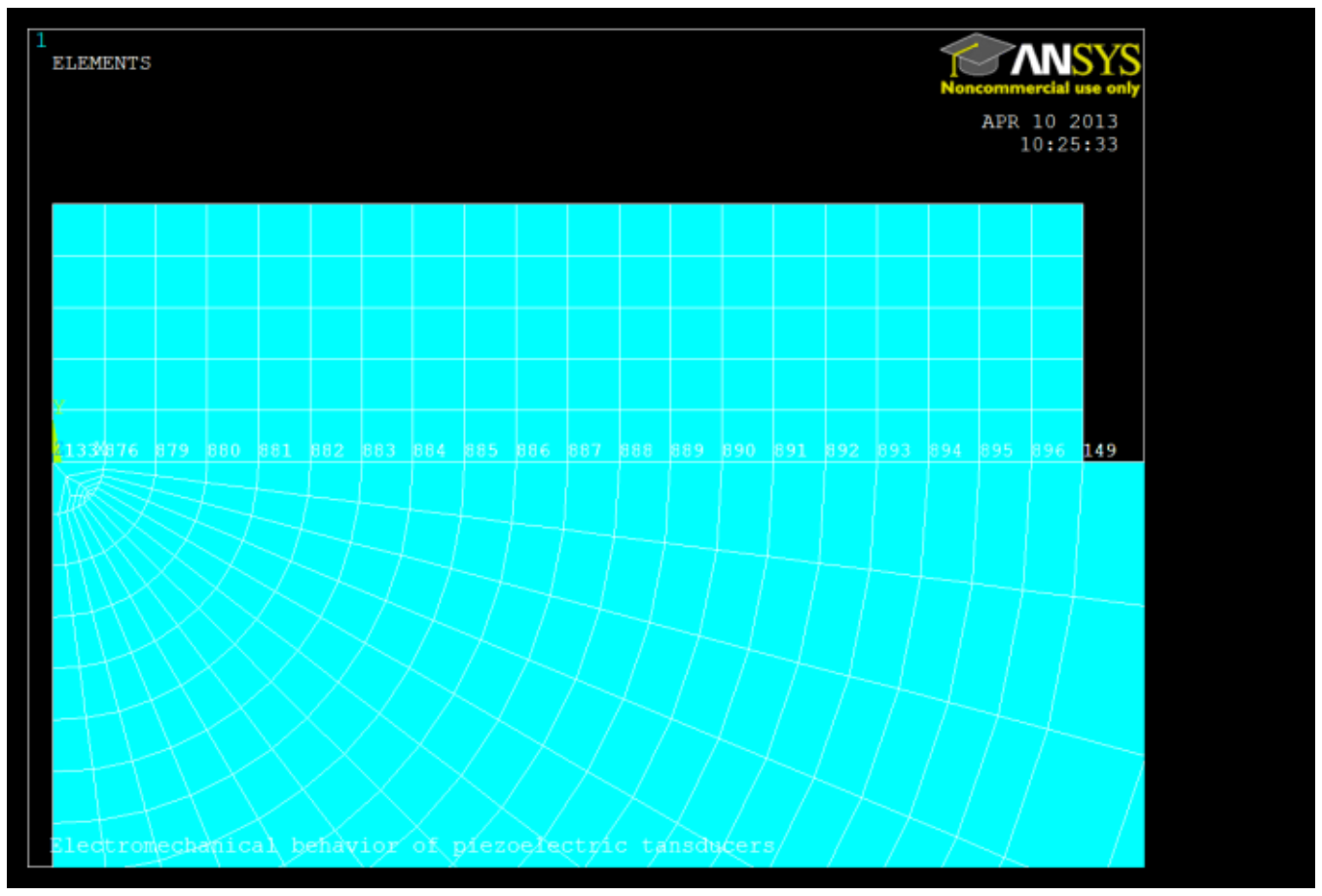

Figure A18. Interface nodal number (involving bonding layer)

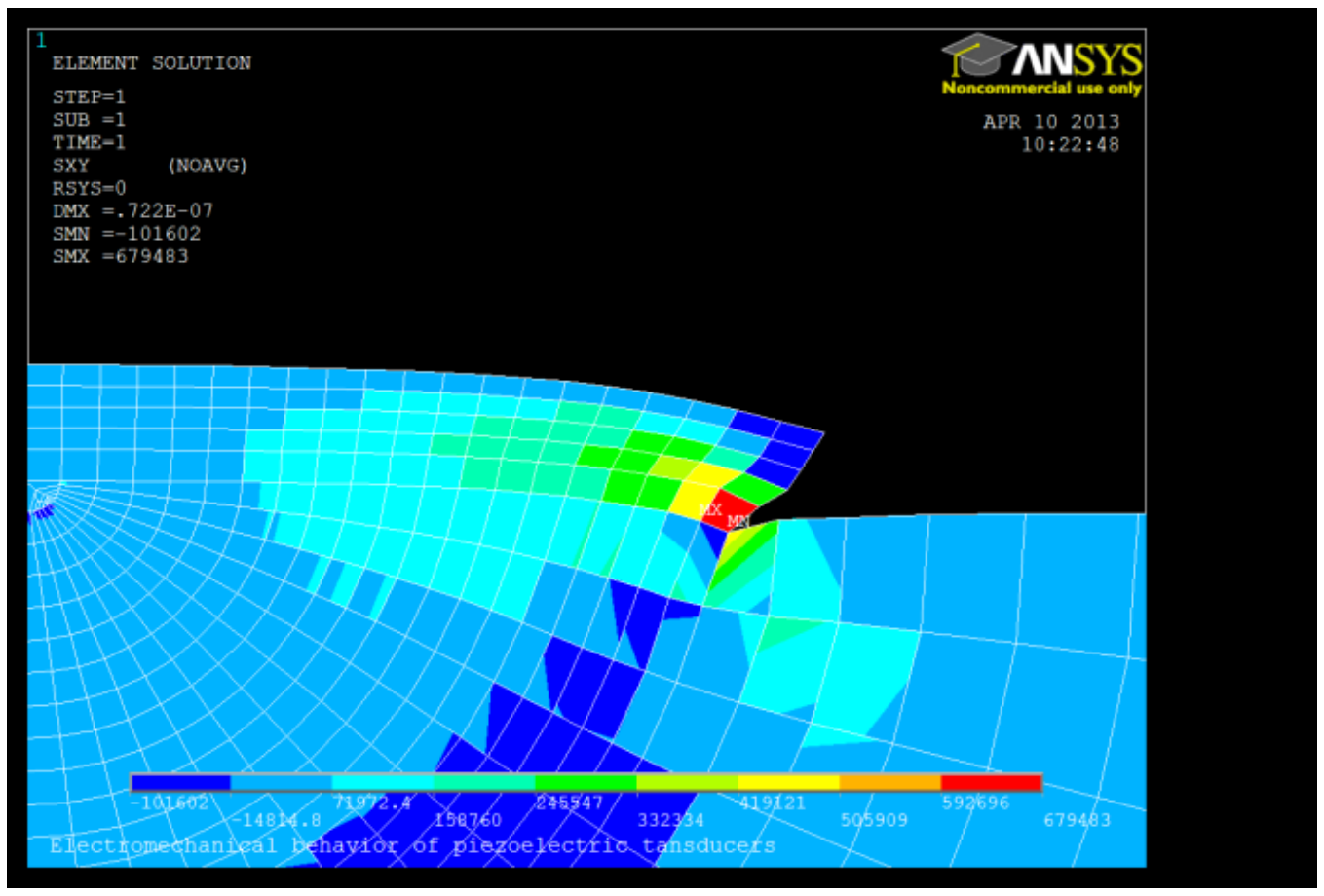

Figure A21. XY shear stress-element solution (involving bonding layer)

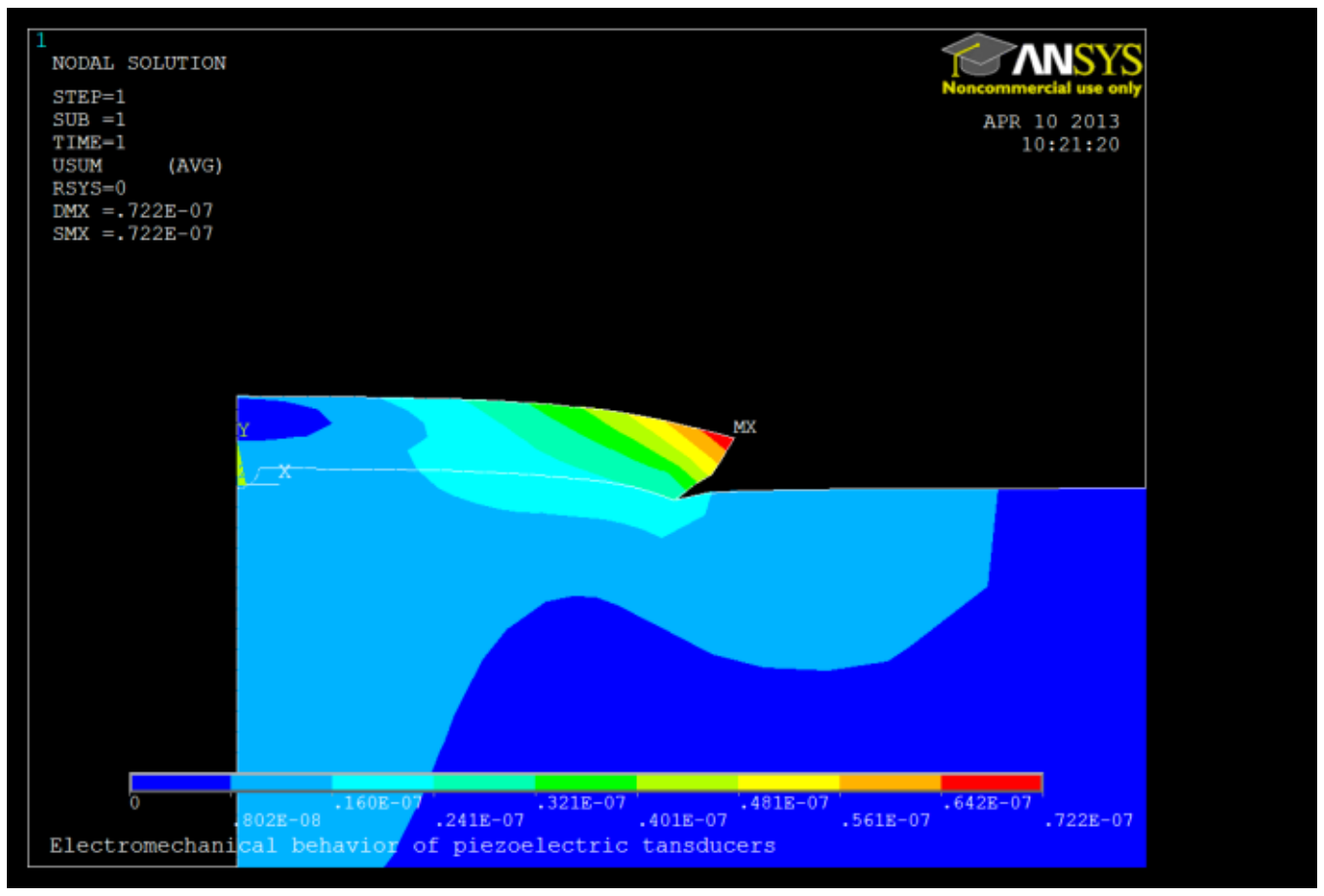

Figure A19. Vector displacement (involving bonding layer)

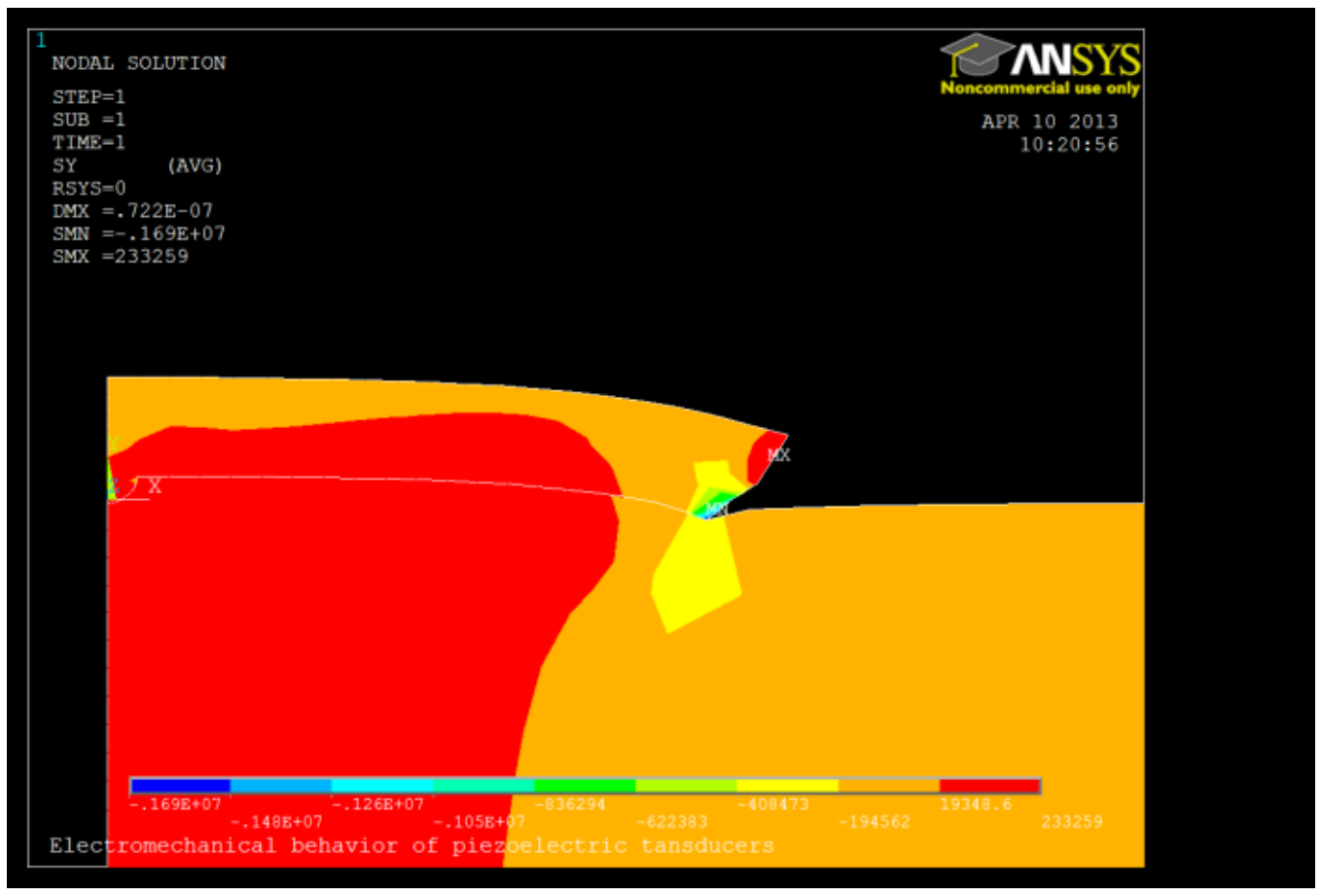

Figure A22. Y stress---nodal solution (involving bonding layer)

Corresponding author: huangcha@ualberta.ca, Department of Mechanic Engineering, University of Alberta, Canada

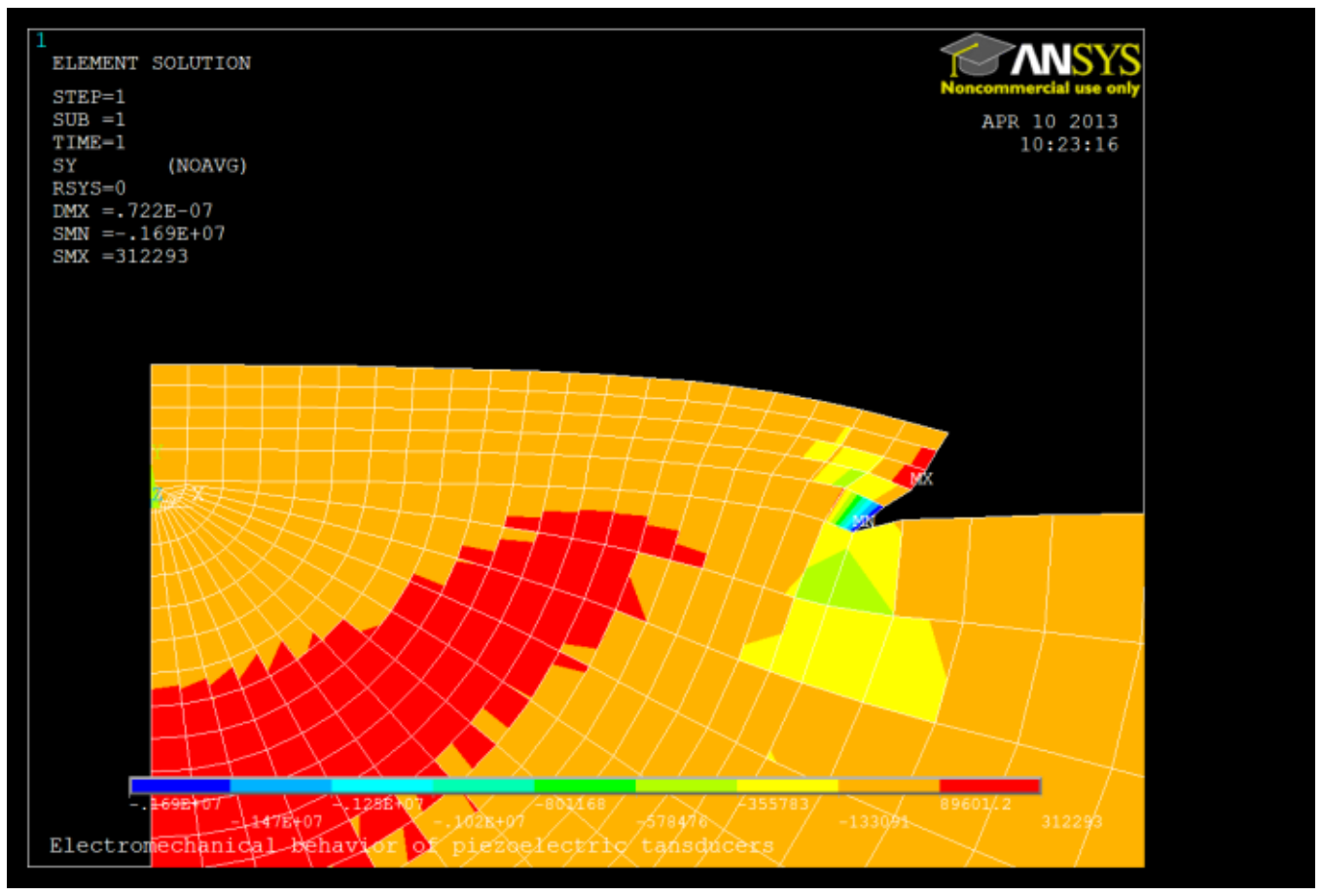

Figure A23. Y stress-element solution (involving bonding layer)

Table A3. Stress result at interface (involving bonding layer)

| x/a | NODE | SX | SY | SXY |
|---|---|---|---|---|
| 0.05 | 876 | 695100 | 20151 | 18845 |
| 0.1 | 879 | 790850 | 42004 | 24957 |
| 0.15 | 880 | 781390 | 32770 | 37216 |
| 0.2 | 881 | 767790 | 37650 | 42962 |
| 0.25 | 882 | 761620 | 39675 | 52926 |
| 0.3 | 883 | 755250 | 43705 | 64017 |
| 0.35 | 884 | 747760 | 48256 | 76002 |
| 0.4 | 885 | 738040 | 53759 | 88722 |
| 0.45 | 886 | 725140 | 59912 | 102120 |
| 0.5 | 887 | 708020 | 66460 | 116141 |
| 0.55 | 888 | 685540 | 72837 | 130723 |
| 0.6 | 889 | 656240 | 78030 | 145903 |
| 0.65 | 890 | 618130 | 80584 | 161957 |
| 0.7 | 891 | 567950 | 77753 | 179805 |
| 0.75 | 892 | 500140 | 67899 | 201241 |
| 0.8 | 893 | 404100 | 43051 | 230350 |
| 0.85 | 894 | 258840 | 17535 | 272237 |
| 0.9 | 895 | -3785 | -74711 | 370000 |
| 0.95 | 896 | -626520 | 13709 | 528000 |

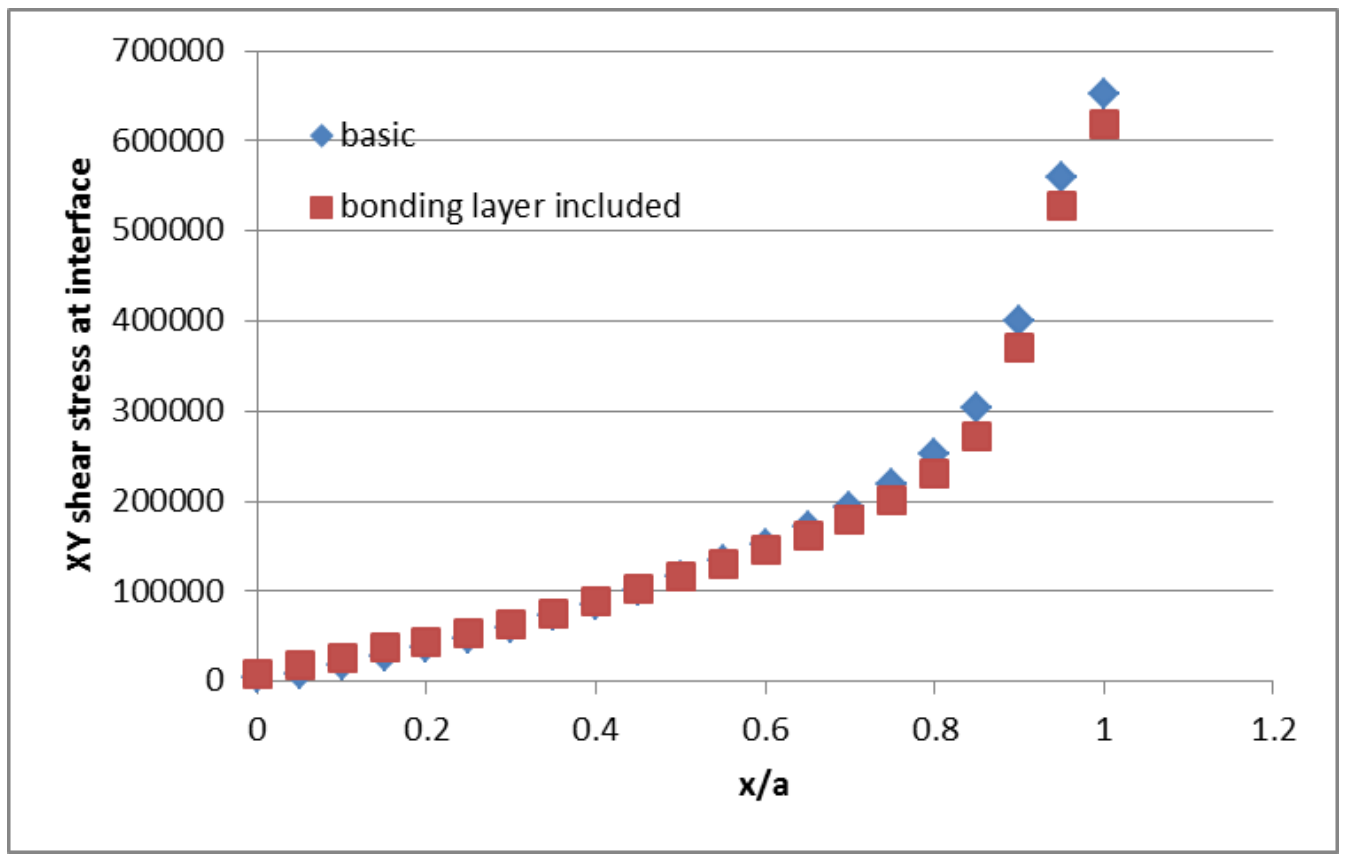

Figure A24. Effect of bonding layer (involving bonding layer)

## 4. Harmonic analysis result

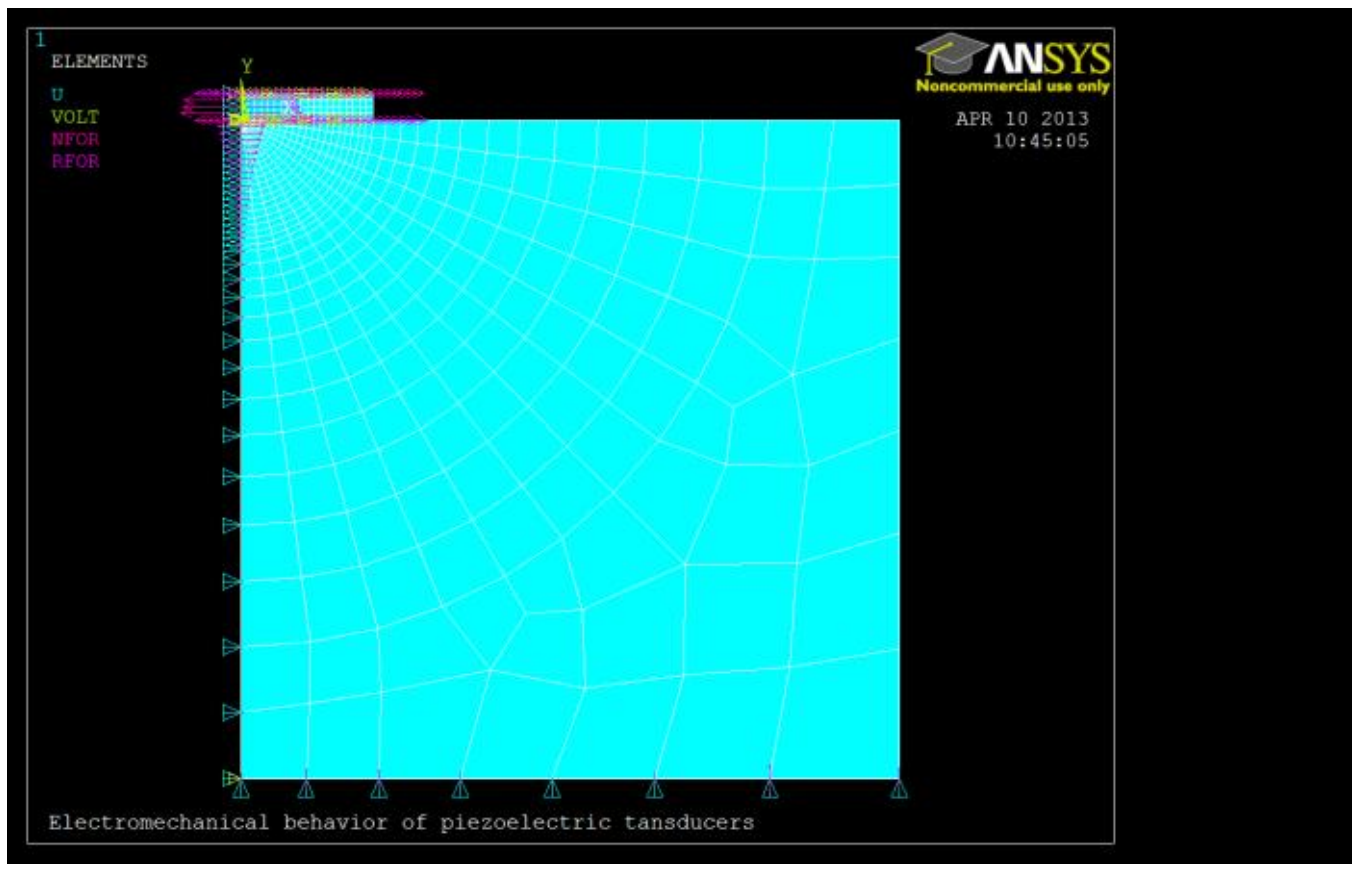

Figure A25. Model and boundary conditions (harmonic analysis)

| Element | Node | Result Item | Minimum | Maximum | X-Axis |
|---|---|---|---|---|---|
| | | Frequency | 1000 | 100000 | ⦿ |
| averaged | 538 | XY Shear stress | -101252 | 14994.3 | ○ |
| averaged | 142 | XY Shear stress | -189003 | 23434.2 | ○ |
| averaged | 153 | XY Shear stress | -364124 | 48254.3 | ○ |
| averaged | 164 | XY Shear stress | -553101 | 74621.5 | ○ |
| averaged | 175 | XY Shear stress | -735098 | 99759.8 | ○ |
| averaged | 186 | XY Shear stress | -905879 | 122983 | ○ |
| averaged | 197 | XY Shear stress | -1.06112e+006 | 143404 | ○ |
| averaged | 208 | XY Shear stress | -1.19712e+006 | 160267 | ○ |
| averaged | 219 | XY Shear stress | -1.31144e+006 | 173029 | ○ |
| averaged | 230 | XY Shear stress | -1.40315e+006 | 181396 | ○ |
| averaged | 241 | XY Shear stress | -1.47303e+006 | 185356 | ○ |
| averaged | 252 | XY Shear stress | -1.5236e+006 | 188968 | ○ |
| averaged | 263 | XY Shear stress | -1.5589e+006 | 194105 | ○ |
| averaged | 274 | XY Shear stress | -1.58391e+006 | 199121 | ○ |
| averaged | 285 | XY Shear stress | -1.60355e+006 | 205178 | ○ |
| averaged | 296 | XY Shear stress | -1.62159e+006 | 213215 | ○ |
| averaged | 307 | XY Shear stress | -1.64049e+006 | 223928 | ○ |
| averaged | 318 | XY Shear stress | -1.66151e+006 | 237757 | ○ |
| averaged | 329 | XY Shear stress | -1.68574e+006 | 255162 | ○ |
| averaged | 340 | XY Shear stress | -1.70581e+006 | 274719 | ○ |
| averaged | 131 | XY Shear stress | -1.71974e+006 | 300799 | ○ |
| averaged | 496 | XY Shear stress | -1.7409e+006 | 351206 | ○ |
| averaged | 485 | XY Shear stress | -1.70301e+006 | 416191 | ○ |
| averaged | 497 | XY Shear stress | -1.58015e+006 | 321060 | ○ |
| averaged | 486 | XY Shear stress | -1.57513e+006 | 379497 | ○ |
| averaged | 475 | XY Shear stress | -1.51163e+006 | 450248 | ○ |
| averaged | 464 | XY Shear stress | -1.36405e+006 | 532554 | ○ |
| averaged | 453 | XY Shear stress | -1.11732e+006 | 624283 | ○ |
| averaged | 442 | XY Shear stress | -786871 | 722622 | ○ |
| averaged | 431 | XY Shear stress | -712625 | 823575 | ○ |
| averaged | 420 | XY Shear stress | -816456 | 921308 | ○ |
| averaged | 409 | XY Shear stress | -769847 | 1.00724e+006 | ○ |
| averaged | 398 | XY Shear stress | -503425 | 1.06534e+006 | ○ |
| averaged | 387 | XY Shear stress | -119215 | 1.07887e+006 | ○ |
| averaged | 374 | XY Shear stress | -420961 | 1.03659e+006 | ○ |

Figure A26. -45 degree line nodal shear stress result (harmonic analysis)

Corresponding author: huangcha@ualberta.ca, Department of Mechanic Engineering, University of Alberta, Canada

Table A4. XY shear stress at interface (harmonic analysis)

| x/a | Freq.=0 | Freq.=1000000 |
|---|---|---|
| 0.05 | 0 | 1000000 |
| 0.1 | 8892 | 32567.54 |
| 0.15 | 18156 | 65914.57 |
| 0.2 | 27727 | 99736.31 |
| 0.25 | 37705 | 133622.4 |
| 0.3 | 48375 | 167290.3 |
| 0.35 | 59846 | 200246.3 |
| 0.4 | 72132 | 232085.3 |
| 0.45 | 85393 | 262299.7 |
| 0.5 | 1.00E+05 | 290266.1 |
| 0.55 | 1.16E+05 | 315366.4 |
| 0.6 | 1.34E+05 | 345541.6 |
| 0.65 | 1.52E+05 | 372300.6 |
| 0.7 | 1.72E+05 | 393145.5 |
| 0.75 | 1.94E+05 | 420951.4 |
| 0.8 | 2.18E+05 | 446203.1 |
| 0.85 | 2.52E+05 | 480582.2 |
| 0.9 | 3.03E+05 | 521012.9 |
| 0.95 | 4.00E+05 | 612972 |

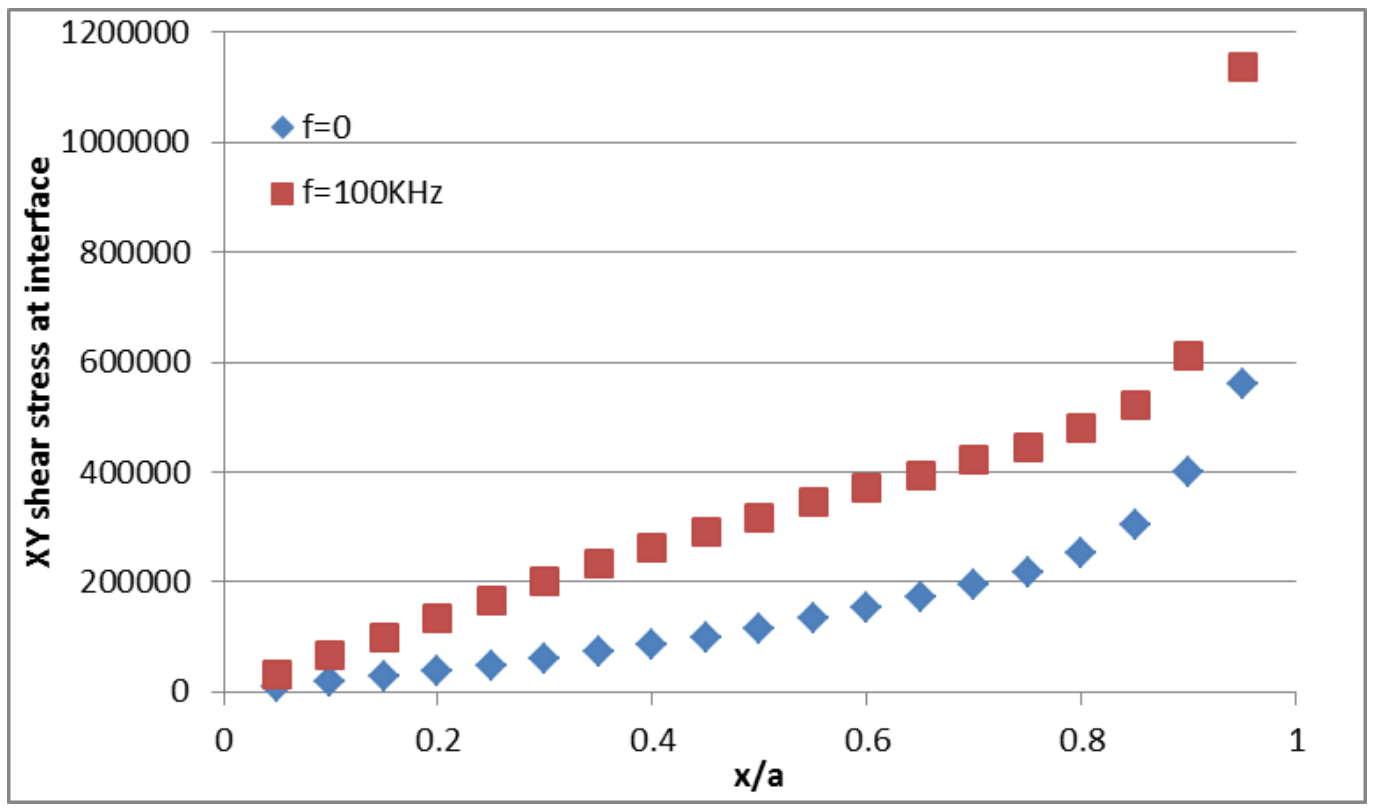

Figure A27 effect of frequency (harmonic analysis)

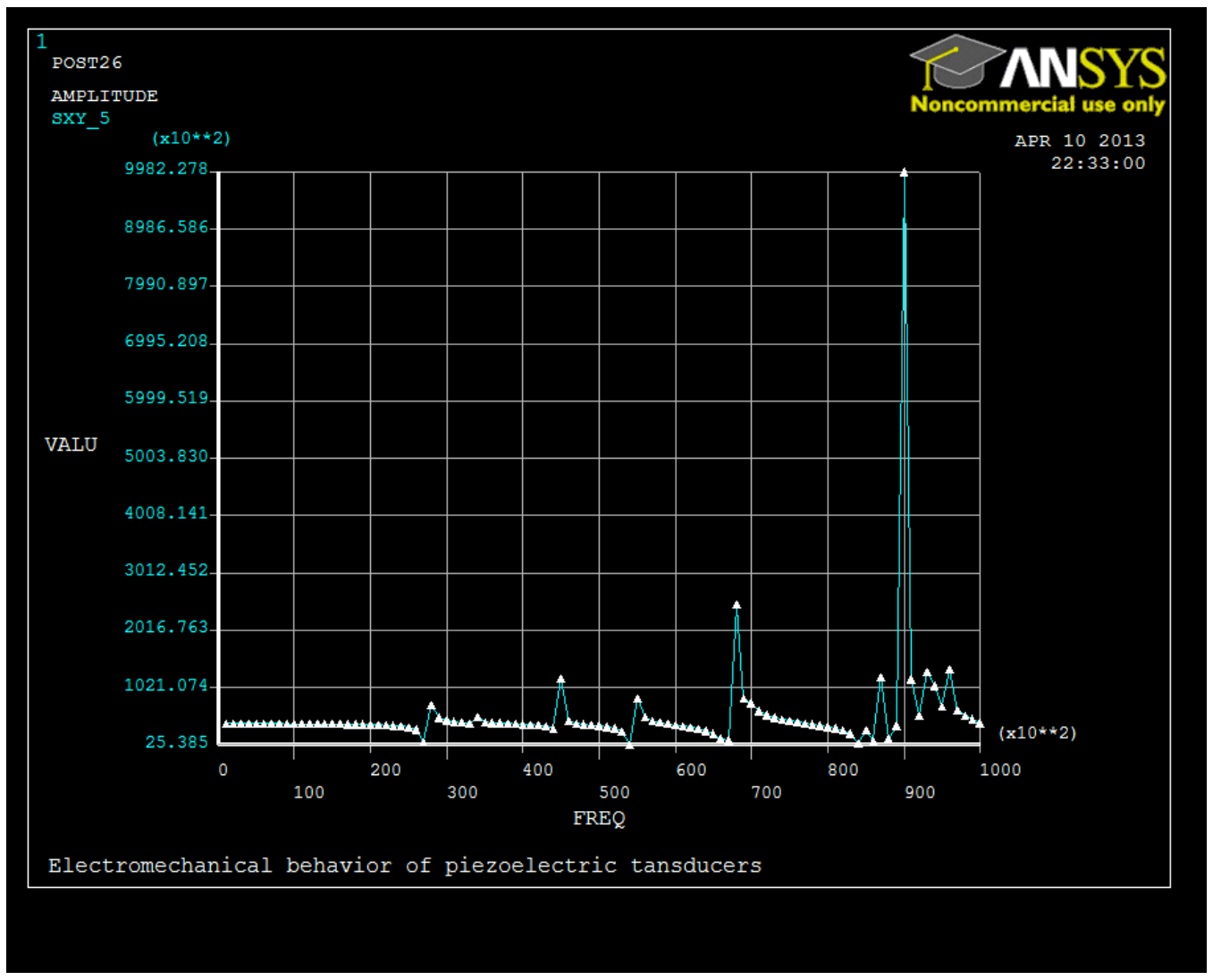

Figure A28. shear&frequency of node 180 near the actuator (harmonic analysis)

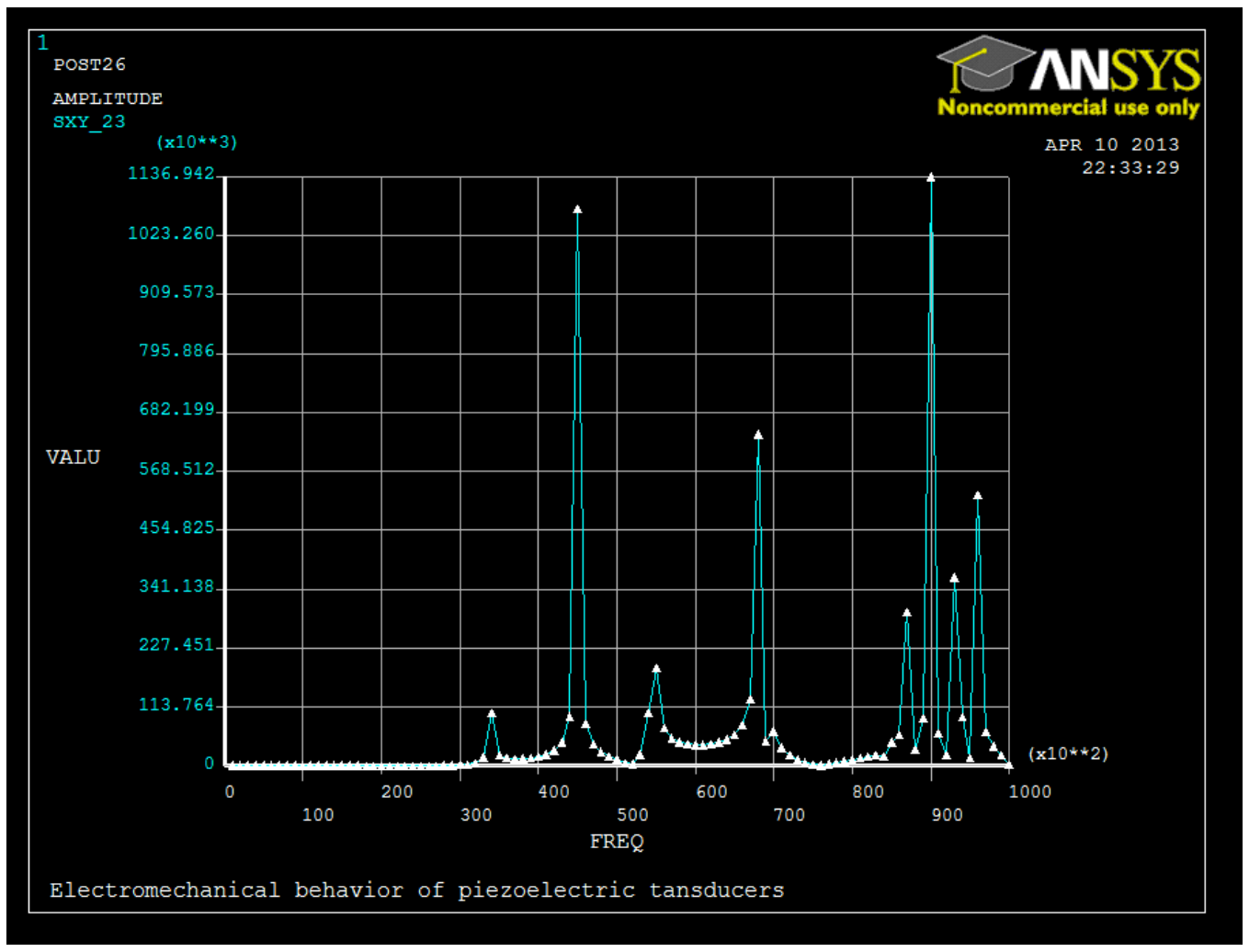

Figure A29. shear&frequency of node429 far from the actuator (harmonic analysis)

Corresponding author: huangcha@ualberta.ca, Department of Mechanic Engineering, University of Alberta, Canada